\documentclass[10pt, two column, twoside]{IEEEtran}

\usepackage{amssymb}
\usepackage{amsmath}
\usepackage[lined,boxed,commentsnumbered, ruled]{algorithm2e}
\usepackage{mathrsfs}

\usepackage{pgfplots}
\usepackage{tikz}
\usetikzlibrary{arrows}
\usepackage{subfigure}
\usepackage{graphicx,booktabs,multirow}

\definecolor{colorhkust}{RGB}{20,43,140}
\definecolor{colortsinghua}{RGB}{116,52,129}
\definecolor{color1}{HTML}{D0B22B}

\newtheorem{lemma}{Lemma}
\newtheorem{theorem}{Theorem}

\newtheorem{proposition}{Proposition}

\newtheorem{remark}{Remark}

\newcommand{\bs}{\boldsymbol}
\newcommand{\tabincell}[2]{\begin{tabular}{@{}#1@{}}#2\end{tabular}}
\newcommand{\rev}{\color{black}}

\begin{document}

\title{Robust Group Sparse Beamforming  for   Multicast Green Cloud-RAN
with Imperfect CSI}
\author{Yuanming~Shi,~\IEEEmembership{Student~Member,~IEEE,}
        Jun~Zhang,~\IEEEmembership{Member,~IEEE,}
        and~Khaled~B.~Letaief,~\IEEEmembership{Fellow,~IEEE}
\thanks{Copyright (c) 2015 IEEE. Personal use of this material is permitted.
However, permission to use this material for any other purposes must be obtained
from the IEEE by sending a request to pubs-permissions@ieee.org. This work
is supported by the Hong Kong Research Grant Council under Grant No. 16200214.}
\thanks{The authors are with the Department of Electronic and Computer Engineering, Hong Kong University of Science and Technology (e-mail: \{yshiac, eejzhang, eekhaled\}@ust.hk).}}

\maketitle

\begin{abstract}
In this paper, we investigate the network power minimization problem for the multicast cloud radio access network (Cloud-RAN) with imperfect channel state information (CSI). The key observation is that network power
minimization can be achieved by adaptively selecting active remote radio heads (RRHs) via controlling the group-sparsity structure of the beamforming vector. However, this yields a non-convex combinatorial optimization problem, for which we propose a three-stage robust group sparse beamforming algorithm. In the first stage, a quadratic variational formulation of the weighted mixed $\ell_1/\ell_2$-norm is proposed to induce the group-sparsity structure in the aggregated beamforming vector, which indicates those RRHs that can be switched off. A perturbed alternating optimization algorithm is then proposed to solve the resultant non-convex group-sparsity inducing optimization problem by exploiting its convex substructures. In the second stage, we propose a \emph{PhaseLift} technique based algorithm to solve the feasibility problem with a given active RRH set, which helps determine the active RRHs. Finally, the semidefinite relaxation (SDR) technique is adopted to determine the robust multicast beamformers.  Simulation results will  demonstrate the convergence of the perturbed alternating optimization algorithm, as well as, the effectiveness of the proposed algorithm to minimize the network power consumption for multicast Cloud-RAN. 
\end{abstract}


\section{Introduction}
\IEEEPARstart{N}{etwork} densification has been recognized as an effective way to meet the exponentially growing mobile data traffic and to accommodate increasingly diversified mobile applications.
Cooperative transmission/reception among multiple base stations is a well-known approach to improve the spectral efficiency and energy efficiency of dense wireless networks \cite{Foschini_2006network,Jun_2009networkedTWC, Gesbert_JSAC10}, which is driving the development of novel collaborative architectures for cellular networks. Cloud radio
access networks (Cloud-RAN) \cite{mobile2011c,Yuanming_TWC2014,Yuanming_WCMLargeCVX}
have recently been proposed as a cost-effective and flexible way to exploit the cooperation gains by moving the baseband units (BBUs) into a single cloud data center, i.e., forming a BBU pool with powerful shared computing resources. As a result, with  efficient hardware utilization at the BBU pool, both the CAPEX (e.g.,
via low-cost site construction) and OPEX (e.g., via centralized cooling)
can be reduced significantly. Furthermore, the conventional base stations are replaced by the light and low-cost remote radio heads (RRHs) with basic functionalities of signal transmission and reception, which are then connected to the BBU pool by high-capacity and low-latency optical fronthaul links.
 The capacity of Cloud-RAN thus can  be significantly improved through network densification and centralized signal processing at the BBU pool. 

However, the new architecture of Cloud-RAN also brings new design and operating challenges, e.g., high-capacity and low-latency requirements for the optical fronthaul links \cite{Shamai_TSP2013}, virtualization techniques for
resource management in the BBU pool  \cite{mobile2011c}, and massive CSI
acquisition for cooperative interference management \cite{Yuaning_ICC2014,Yuanming_TSP14SCB}. In particular, energy efficiency is an important aspect for operating such a dense wireless network, and it is among the major design objectives for 5G networks \cite{Jeff_JSAC5G}. Conventionally, the energy efficiency oriented design only takes into account the transmit power \cite{WeiYu_WC10}  and the circuit power \cite{Goldsmith_JSAC2004energy}  at the base stations. Nevertheless, in such dense collaborative networks as Cloud-RAN, a holistic  view is needed when measuring network power consumption, which should also include the power consumption of the additional optical fronthaul links \cite{Yuanming_TWC2014}. Observing that the mobile data traffic would vary temporally and spatially, it was proposed in [5] to adaptively switch off some fronthaul links and the corresponding RRHs to minimize  the network power consumption, which is achieved by a new beamforming technique, called \emph{group sparse beamforming}.

The effectiveness of group sparse beamforming has been demonstrated in [5], but with certain limitations in the network model, e.g., perfect CSI is assumed at the BBU pool, and only unicast services are considered. In practice, inevitably there will be uncertainty in the available CSI, originating from various sources, e.g., limited feedback \cite{love2008overview}, channel estimation errors \cite{Jindal_TC2010unified},
partial CSI acquisition \cite{Yuaning_ICC2014, Yuanming_TSP14SCB} and delay in the obtained
CSI \cite{Tse_TIT2012completely,Jun_2009mode}. In terms of transmission services from the RRHs, it has been well recognized that the physical layer integration
technique \cite{Schaefer_SPM2014Physcial}
can effectively  improve the network performance. In particular, the RRHs
should not
only transmit data to individual users \cite{shamai2006capacity} (i.e., broadcast/unicast
services) but also integrate additional multicast services \cite{Luo_2008quality}, where the RRHs transmit a common message in such a way that all the MUs in the same  group can decode it.
Such multigroup multicast transmission is promising to
provide high capacity services and content-aware applications in next generation
wireless networks. For instance, with physical layer caching for wireless
video delivery \cite{Caire_CM2013}, it is common that multiple users are interested in  the same video stream, which creates multicast groups.

In this paper, we will thus focus on the design of green Cloud-RAN by jointly minimizing the RRH power consumption and transport link power consumption, considering the practical scenarios with imperfect CSI and multigroup multicast services. We adopt the robust optimization approach to address the CSI uncertainty, such that the QoS requirements are satisfied for \emph{any} realization of the uncertainty in a predefined set \cite{bertsimas2011theory}. The unique challenges of the network power minimization problem arise from both the infinite number of the  non-convex quadratic QoS constraints (due to the robust design criteria and multicast transmission) and the combinatorial composite objective function (due to the consideration of both the relative fronthaul link power consumption and the RRH transmit power consumption).

\subsection{Related Works}
\subsubsection{Robust Multicast Beamforming}
Although the integration of multicast, individual services and cooperative
transmission can significantly improve the capacity of wireless networks \cite{Schaefer_SPM2014Physcial}, it will
bring significant challenges from both the
information theoretic \cite{Shamai_TIT2009multicast} and signal processing
perspectives \cite{Luo_2008quality,Jorswieck_TSP2011optimal}. In particular,
the physical-layer multicast beamforming problem is in general NP-hard
 due to the non-convex quadratic QoS constraints \cite{Luo_2008quality}. Furthermore, to address the CSI uncertainty, one may either adopt the stochastic optimization
formulation \cite{shapiro2009lectures} or the robust optimization formulation
\cite{ben2009robust}. However, the stochastic
optimization formulations often yield highly intractable problems, e.g., the
stochastic coordinated beamforming problem based on the chance constrained
programming \cite{Yuanming_TSP14SCB}. The worst-case based robust optimization,
on the other hand,  has the advantage of computational tractability
\cite{bertsimas2011theory}. Although the original robust and/or multicast beamforming design
problems may be non-convex due to the infinite number of non-convex quadratic QoS constraints \cite{shen2012distributed}, the convex optimization based SDR technique \cite{Z.Q.Luo_SPM2010} with S-lemma \cite{boyd2004convex}
has recently been  applied to   provide a principled way to develop polynomial
time complexity algorithms to find an approximate solution \cite{Z.Q.Luo_2012robust}. 

However, we cannot directly apply such SDR technique to solve the network power  minimization problem due to the non-convex combinatorial composite objective function, which represents the network power consumption.

\subsubsection{Group Sparse Beamforming}

The convex sparsity-inducing penalty  approach \cite{Bach_ML2011} has recently been widely used to develop polynomial time complexity algorithms for  the mixed combinatorial optimization problems in wireless networks, e.g., joint base station clustering and transmit beamforming \cite{Z.Q.Luo_JSAC2013}, joint antenna \cite{Mehanna_SP2013} or RRH \cite{Yuanming_TWC2014} selection and transmit beamforming. The main idea of this approach is that the sparsity pattern of the beamforming vector, which can be induced by minimizing a sparsity penalty function (e.g., the mixed $\ell_1/\ell_2$-norm minimization can induce the group-sparsity), can provide guidelines for, e.g., antenna selection \cite{Mehanna_SP2013}, where the antennas with smaller beamforming coefficients (measured by the $\ell_\infty$-norm) have a higher priority to be switched off. However, most works only consider the ideal scenario (e.g., perfect CSI and broadcast services \cite{Yuanming_TWC2014}), which usually yield convex constraints (e.g., second-order cone constraints \cite{Yuanming_TWC2014}).
 
Unfortunately, we cannot directly adopt the \emph{non-smooth} weighted mixed  $\ell_1/\ell_2$-norm developed in \cite{Yuanming_TWC2014} to induce the group-sparsity for the robust multicast beamforming vector. This is because the resultant group-sparsity inducing optimization problem will be highly intractable, due to the non-smooth sparsity-inducing objective function and the infinite number of non-convex quadratic QoS constraints.    

{\rev{Based on  above discussion and  in contrast to the previous work \cite{Yuanming_TWC2014} on  group sparse beamforming   with a non-convex combinatorial composite objective function but convex QoS constraints in the unicast Cloud-RAN, we need to address the following coupled challenges in order to solve the network power minimization problem for multicast green Cloud-RAN with imperfect CSI}}:
\begin{itemize}
\item An infinite number of non-convex quadratic QoS constraints;
\item The combinatorial composite objective function. 
\end{itemize}  
Thus, to apply the computationally efficient group sparse beamforming approach \cite{Yuanming_TWC2014}  to more practical scenarios, unique challenges arise. We need to redesign the group-sparsity inducing norm, and then deal with the non-convex group-sparsity inducing optimization problem with an infinite number of non-convex quadratic QoS constraints. We should also develop efficient algorithms for non-convex feasibility problems for the adaptive RRH selection, and for non-convex robust multicast beamforming design after determining the active RRHs.

\subsection{Contributions}
In this paper,  we provide a  convex relaxation based robust group sparse beamforming framework for network power minimization in multicast Cloud-RAN with imperfect CSI. 
The major contributions are summarized as follows:
\begin{enumerate}
\item A group sparse beamforming formulation  is proposed to minimize the network power consumption for Cloud-RAN. It will simultaneously control the group-sparsity structure and the magnitude of the
beamforming coefficients, thereby minimizing the
relative fronthaul link power consumption and the transmit power consumption,
respectively. The group sparse beamforming modeling framework lays the foundation for developing the three-stage robust group sparse beamforming algorithm based on the convex relaxation.    

\item In the first stage, a  novel quadratic variational formulation of the weighted mixed $\ell_1/\ell_2$-norm is proposed to induce the group-sparsity structure for the robust multicast beamforming vector, thereby guiding the RRH selection. The main motivation for such a quadratic form formulation is to make the group-sparsity inducing penalty function  compatible with the quadratic QoS constraints. Based on the SDR technique, a perturbed alternating optimization algorithm  with convergence guarantee is then proposed to solve the resultant non-convex quadratic form group-sparsity inducing optimization problem by exploiting its convex substructures. 
\item In the second stage, a \emph{PhaseLift} approach based algorithm is proposed to solve the non-convex feasibility problems, based on which the active RRHs can be determined with a binary search. Finally, the SDR technique is adopted to solve the non-convex robust multicast beamforming optimization problem to determine the  transmit beamformers for the active RRHs. 
\item Simulation results will demonstrate the effectiveness of the proposed robust group sparse beamforming algorithm to minimize the network power consumption.        
\end{enumerate}

\subsection{Organization}
The remainder of the paper is organized as follows. Section {\ref{sysprob}}
presents the system model and problem formulation, followed by the problem
analysis. In Section {\ref{GSBFModel}}, the  group sparse beamforming modeling framework is proposed to formulate the the network power minimization problem. The semidefinite programming (SDP) based robust group sparse beamforming algorithm is developed in Section {\ref{sdrfra}}.
Simulation results will be illustrated in Section {\ref{simu}}.
Finally, conclusions and discussions are presented in Section {\ref{confw}}.

\section{System Model and Problem Formulation}
\label{sysprob}
\subsection{System Model}
\begin{figure}[t]
  \centering
  \includegraphics[width=0.95\columnwidth]{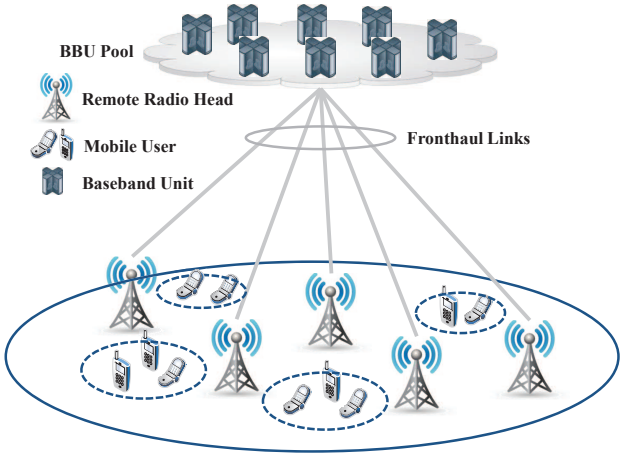}
 \caption{The architecture of the multicast Cloud-RAN, in which, all the RRHs are connected to a BBU pool through high-capacity and low-latency
optical fronthaul links. All the MUs in the same dashed circle form a multicast group and request the same message.}
  \label{cran}
 \end{figure}
Consider a multicast Cloud-RAN with $L$ RRHs and $K$ single-antenna mobile users (MUs), where the $l$-th RRH
is equipped with $N_{l}$ antennas, as shown in Fig. {\ref{cran}}.
 The centralized signal processing is performed at the baseband unit (BBU)
pool \cite{ mobile2011c,Yuanming_TWC2014}.  Define $\mathcal{S}=\{1,\dots, K\}$ as the set of all the MUs and $\mathcal{L}=\{1,\dots, L\}$ as the set of all the RRHs. We focus on the downlink transmission, for which the signal  processing is
more challenging. Assume that there are $M$ ($1\le M\le K$) multicast groups, i.e., $\{\mathcal{G}_1,\dots,
\mathcal{G}_M\}$, where $\mathcal{G}_m$ is the set of MUs in the multicast
group $m$ with $1\le m\le M$.  Let $\mathcal{M}=\{1,\dots, M\}$ be the set of the multicast groups. Each MU only belongs to a single multicast
group, i.e., $\mathcal{G}_i\cap\mathcal{G}_j=\emptyset$ such that $\cup_{i}\mathcal{G}_i=\mathcal{S}$
and $\sum_{i}|\mathcal{G}_i|=K$.

 Let ${\bf{v}}_{lm}\in\mathbb{C}^{N_{l}}$ be the transmit beamforming vector
from the $l$-th RRH to the $k$-th MU in group $\mathcal{G}_m$. The  encoded
transmission information symbol of the multicast group $m$ is denoted as $s_m\in\mathbb{C}$ with $\mathbb{E}[|s_m|^2]=1$.
The channel propagation between  MU $k$ and RRH $l$ is denoted as ${\bf{h}}_{kl}\in\mathbb{C}^{N_{l}}$.
Therefore, the received signal $y_{k,m}\in\mathbb{C}$ at MU $k$ in the multicast group $m$
is given
by
\setlength\arraycolsep{1.2pt}
\begin{eqnarray}
\!\!\!\!\!y_{k, m}=\sum_{l=1}^{L}{\bf{h}}_{kl}^{\sf{H}}{\bf{v}}_{lm}s_{m}+\sum_{i\ne
m}\sum_{l=1}^{L}{\bf{h}}_{kl}^{\sf{H}}{\bf{v}}_{li}s_{i}+n_{k}, \forall k\in\mathcal{G}_m,
\end{eqnarray}  
where $n_{k}\sim\mathcal{CN}(0, \sigma_{k}^2)$ is the additive Gaussian noise
at MU $k$. We assume that $s_{m}$'s
and $n_{k}$'s are mutually independent and all the MUs apply single user
detection. The signal-to-interference-plus-noise ratio (SINR) for MU $k\in\mathcal{G}_m$
is given by
\begin{eqnarray}
{{\Gamma}}_{k,m}={{|{\bf{h}}_{k}^{\sf{H}}{\bf{v}}_{m}|^2}\over{\sum_{i\ne
m}|{\bf{h}}_{k}^{\sf{H}}{\bf{v}}_{i}|^2+\sigma_{k}^2}}, \forall k\in\mathcal{G}_m,
\end{eqnarray}
where ${\bf{h}}_{k}\triangleq [{\bf{h}}_{k1}^{T},\dots, {\bf{h}}_{kL}^{T}]^{T}\in\mathbb{C}^{N}$
with $N=\sum_{l=1}^{L}N_{l}$, and ${\bf{v}}_{m}\triangleq[{\bf{v}}_{1m}^{T},
{\bf{v}}_{2m}^{T},\dots, {\bf{v}}_{Lm}^{T}]^{T}\in\mathbb{C}^{N}$ is the
aggregative beamforming vector for the multicast group $m$ from all the RRHs.  The transmit signal at RRH $l$ is given by
\begin{eqnarray}
{\bf{x}}_l=\sum_{m=1}^{M}{\bf{v}}_{lm}s_m, \forall l.
\end{eqnarray}
Each RRH has its own transmit power constraint, i.e., 
\begin{eqnarray}
\sum_{m=1}^{M}\|{\bf{v}}_{lm}\|_2^2\le P_l, \forall l,
\end{eqnarray}
where $P_l>0$ is the maximum transmit power of RRH $l$.

\subsection{Problem Formulation}

\subsubsection{Imperfect CSI}
In practice, the CSI at the BBU pool will be imperfect, which may originate
from a variety of sources. For instance, in frequency-division duplex (FDD) systems, the CSI
imperfection may originate from downlink training based channel estimation
\cite{Jindal_TC2010unified} and uplink limited feedback \cite{love2008overview}.
It could also be due to the hardware deficiencies, partial CSI acquisition \cite{Yuaning_ICC2014, Yuanming_TSP14SCB} and delays in CSI acquisition
\cite{Tse_TIT2012completely,Jun_2009mode}.
In this paper, we adopt the following additive error model \cite{shen2012distributed,Emil_TSP2012,Bjornson_TCIT2013}
to model the channel imperfection from all the RRHs to MU $k$, i.e., 
\begin{eqnarray}
{\bf{h}}_{k}=\hat{\bf{h}}_{k}+{\bf{e}}_{k}, \forall k, 
\end{eqnarray}
where $\hat{\bf{h}}_{k}$ is the estimated channel vector and ${\bf{e}}_{k}$ is the estimation error vector. {\rev{There are mainly two ways to model the CSI uncertainty: one is the stochastic modeling based on the probabilistic description, and the other is the deterministic and set-based modeling. However, the stochastic CSI uncertainty modeling will yield probabilistic QoS constraints. The resulting chance constrained programming problems are highly intractable in general \cite{Yuanming_TSP14SCB}. Therefore, to seek a computationally tractable formulation,}} we further assume that the error vectors satisfy
the following elliptic model \cite{shen2012distributed,Emil_TSP2012,Bjornson_TCIT2013}:
\begin{eqnarray}
\label{ellp}
{\bf{e}}_{k}^{\sf{H}}{\bf{\Theta}}_{k}{\bf{e}}_{k}\le1, \forall k,
\end{eqnarray}
where ${\boldsymbol{\Theta}}_{k}\in\mathbb{H}^{N\times N}$ with ${\boldsymbol{\Theta}}_{k}\succeq
{\bf{0}}$ is the shape of the ellipsoid. {\rev{This model is motivated by viewing the channel estimation as the main source of CSI uncertainty \cite[Section 4.1]{Bjornson_TCIT2013}.}}
 
\subsubsection{Network Power Consumption}
In Cloud-RAN, it is vital to minimize the network power consumption,
consisting of RRH transmit power and relative fronthaul network power 
\cite{Yuanming_TWC2014}, in order to design a green wireless network. RRH selection will be adopted for this purpose. Specifically, let $\mathcal{A}$
be the set of active RRHs, the network power consumption is given by
\begin{eqnarray}
\label{npc}
p(\mathcal{A})=\sum_{l\in\mathcal{A}}P_l^{c}+\sum_{l\in\mathcal{A}}\sum_{m=1}^{M}{1\over{\eta_{l}}}\|{\bf{v}}_{lm}\|_2^2,
\end{eqnarray}  
where $P_{l}^{c}\ge 0$ is the relative fronthaul link power consumption \cite{Yuanming_TWC2014} (i.e., the static power saving when both the fronthaul link and the corresponding RRH are switched off) and $\eta_{l}>0$ is the drain inefficiency coefficient of the radio frequency power amplifier. The typical values  are $P_{l}^c=5.6 W$ and $\eta_l=25\%$
\cite{Yuanming_TWC2014}, respectively. 

Given the QoS thresholds ${\bs{\gamma}}=(\gamma_1,\dots, \gamma_K)$, in
this paper, we aim at minimizing the network power consumption while guaranteeing the worst-case QoS requirements in the presence of CSI uncertainty and the per-RRH power constraints, i.e., we will consider the following non-convex mixed combinatorial  robust multicast beamforming optimization problem,
\begin{eqnarray}
\mathscr{P}:
\label{comob}
\mathop {\textrm{minimize}}_{{\bf{v}}, \mathcal{A}, \mathcal{Z}}&&\sum_{l\in\mathcal{A}}P_l^{c}+\sum_{l\in\mathcal{A}}\sum_{m=1}^{M}{1\over{\eta_{l}}}\|{\bf{v}}_{lm}\|_2^2\\
\label{robustc1}
\textrm{subject to}&&\sum_{m=1}^{M}\|{\bf{v}}_{lm}\|_2^2\le P_l, \forall l\in\mathcal{A}
\\
\label{pz}
&&\sum_{m=1}^{M}\|{\bf{v}}_{lm}\|_2^2=0, \forall l\in\mathcal{Z}\\
\label{qos1}
&&{{|(\hat{\bf{h}}_{k}+{\bf{e}}_{k})^{\sf{H}}{\bf{v}}_{m}|^2}\over{\sum_{i\ne
m}|(\hat{\bf{h}}_{k}+{\bf{e}}_{k})^{\sf{H}}{\bf{v}}_{i}|^2+\sigma_{k}^2}}\ge
\gamma_k  \\
\label{qos2}
&&{\bf{e}}_{k}^{\sf{H}}{\bf{\Theta}}_{k}{\bf{e}}_{k}\le1, \forall k\in\mathcal{G}_m,
m\in\mathcal{M},
\end{eqnarray}
where $\mathcal{Z}$ is the set of inactive RRHs such that $\mathcal{A}\cup\mathcal{Z}=\mathcal{L}$
and ${\bf{v}}=[{\bf{v}}_{lm}]$ is the aggregated beamforming vector from all the RRHs to all the MUs. The constraints in (\ref{pz}) indicate that
the transmit powers of the inactive RRHs are enforced to be zero. That is, the beamforming coefficients at the inactive RRHs are set to be zero simultaneously. Constraints (\ref{qos1}) and  (\ref{qos2}) indicate that  all the QoS requirements in (\ref{qos1}) should be satisfied for \emph{all} realizations of the errors ${\bf{e}}_k$'s within the feasible set formed by the constraint (\ref{qos2}).  

The network power minimization problem $\mathscr{P}$ imposes the following challenges:
\begin{enumerate}
\item For a given set of CSI error vectors ${\bf{e}}_k$'s, the corresponding network power minimization problem is highly intractable, due to the combinatorial composite objective function (\ref{comob}) and the non-convex quadratic  constraints (\ref{pz}) and (\ref{qos1}). 
\item There are an infinite number of non-convex quadratic QoS constraints due to the worst-case design criterion.  
\end{enumerate} 

To efficiently address the above unique challenges in a unified fashion, in this paper, we will propose a systematic convex relaxation approach based on SDP optimization to solve problem ${\mathscr{P}}$. In particular, the combinatorial challenge will be addressed by the sparsity-inducing penalty  approach in Section {\ref{gspa}}, based on the quadratic variational formulation for the weighted mixed $\ell_1/\ell_2$-norm. The convex optimization technique based on PhaseLift, SDR and S-lemma will be adopted to cope with the infinite number of non-convex quadratic constraints  in Sections {\ref{rrhsel} and \ref{sdr}}. 

In the next subsection, we will provide a detailed analysis of problem $\mathscr{P}$. In particular, the connections with the formulations in existing literatures will be discussed, which will reveal the generality of the formulation $\mathscr{P}$ for practical design problems in Cloud-RAN.

\subsection{Problem Analysis} 
While problem $\mathscr{P}$ incorporates most of the practical elements 
in Cloud-RAN, i.e., imperfect CSI and multigroup multicast transmission, it raises unique challenges compared with the existing
works. Following is a list of key aspects of the difficulty of problem $\mathscr{P}$, accompanied with potential solutions.
\begin{itemize}
\item {\emph{Robust Beamforming Design}}: Suppose that all the RRHs are active,
i.e., $\mathcal{A}=\mathcal{L}$, with broadcast/unicast transmission, i.e., $|\mathcal{G}_m|=1,
\forall m$ and $M=K$. Then problem $\mathscr{P}$ reduces to the conventional worst-case non-convex robust beamforming
design problems \cite{shen2012distributed,Emil_TSP2012}. For this special case, the SDR technique  \cite{Z.Q.Luo_SPM2010}
combined with the S-lemma \cite{boyd2004convex} is proven to be powerful to find good approximation
solutions to such problems. 
\item {\emph{Multicast Beamforming Design}}: Physical-layer multicast beamforming design problems \cite{Luo_2008quality}
prove to be non-convex quadratically constrained problems (QCQP)
\cite{boyd2004convex}, even with perfect CSI and all the RRHs active.
Again, the SDR technique can relax this problem to a convex one, yielding efficient approximation solutions. 
\item {\emph{Quadratically Constrained Feasibility Problem}}: Suppose that the inactive RRH
set $\mathcal{Z}$ with $|\mathcal{Z}|>0$ is fixed, then we have the quadratic
equation constraints (\ref{pz})  in problem $\mathscr{P}$. {\emph{PhaseLift}} \cite{candes_2012phaseretrieval}
is a convex programming technique to relax the non-convex feasibility problem with such quadratic  equation constraints to a convex one
by lifting the problem to higher dimensions and relaxing the rank-one constraints by the convex surrogates, i.e., the trace norms or nuclear norms.
    
\item {\emph{Non-convex Mixed-integer Nonlinear  Programming Optimization Problem}}: Problem
$\mathscr{P}$ can be easily reformulated as a mixed-integer non-linear programming
(MINLP) problem as shown in \cite{Yuanming_TWC2014}. However, the MINLP problem has exponential complexity \cite{leyffer_2012mixed}. Therefore, such a reformulation cannot bring algorithmic design advantages. One thus has to resort to some global optimization techniques \cite{Cheng_SP2013, Yuanming_Globecom2013} (e.g, branch-and-bound method) or greedy algorithms \cite{Yuanming_TWC2014}. Instead, the group-sparsity inducing penalty approach has recently received
enormous attention to seek effective convex relaxation for the MINLP problems, e.g., for jointly designing transmit beamformers and selecting bases stations \cite{Z.Q.Luo_JSAC2013}, transmit antennas \cite{Mehanna_SP2013}, or RRHs \cite{Yuanming_TWC2014}. However, with multicast transmission and imperfect CSI, we cannot directly adopt the  group-sparsity inducing penalty developed in \cite{Yuanming_TWC2014} with the weighted mixed $\ell_1/\ell_2$-norm, as we have seen that we need to lift the problem $\mathscr{P}$ to higher dimensions to cope with the non-convexity of the robust multicast beamforming problem. This requires to develop a new group-sparsity inducing penalty function, which needs to be compatible with quadratic forms, as the beamforming coefficients will be lifted to higher dimensions.     

\end{itemize} 

The above discussions show that problem $\mathscr{P}$ cannot be directly solved by existing methods. Thus, we will propose a new robust group sparse beamforming algorithm in this paper, to solve the highly intractable problem $\mathscr{P}$.  Specifically, in Section {\ref{GSBFModel}}, we will propose a group sparse beamforming modeling framework to reformulate the original problem $\mathscr{P}$. The algorithmic advantages of working with the group sparse beamforming formulation will be revealed in Section {\ref{sdrfra}}, where a robust group sparse beamforming algorithm will be developed.

\section{A Group Sparse Beamforming modeling framework}
\label{GSBFModel}
In this section, we propose a group sparse beamforming modeling framework to reformulate the network power minimization problem $\mathscr{P}$ by controlling the group-sparsity structure and the magnitude of the beamforming coefficients simultaneously. The main advantage of such a modeling framework is the capability of enabling polynomial time complexity algorithm design via convex relaxation.

\subsection{Network Power Consumption Modeling}
We observe that the network power consumption (\ref{npc}) can  be modeled by a composite function parameterized by the  aggregative beamforming coefficients ${\bf{v}}\in\mathbb{C}^{NM}$, which can be written
as a partition
\begin{eqnarray}
\label{vector1}
{\bf{v}}=[\underbrace{{\bf{v}}_{11}^{T},\dots,{\bf{v}}_{1M}^{T}}_{\tilde{\bf{v}}_{1}^{T}},\dots,\underbrace{{\bf{v}}_{L1}^{T},\dots,{\bf{v}}_{LM}^{T}}_{\tilde{\bf{v}}_{L}^{T}}]^{T},
\end{eqnarray} 
where all the coefficients in a given vector $\tilde{\bf{v}}_{l}=[{\bf{v}}_{l1}^{T},\dots,{\bf{v}}_{lM}^{T}]^{T}\in\mathbb{C}^{MN_{l}}$
form a beamforming coefficient group. Specifically, observe that the optimal
aggregative beamforming vector ${\bf{v}}$ in problem
$\mathscr{P}$ should have the group-sparsity structure. That is, when the
RRH $l$ is switched off, the corresponding coefficients in the beamforming
vector $\tilde{\bf{v}}_{l}$ will be set to zero simultaneously. Overall
there may be multiple RRHs being switched off and the corresponding beamforming
vectors will be set to zero, yielding a group-sparsity structure in the beamforming vector $\bf{v}$. 

Define the support of the beamforming vector
${\bf{v}}$ as
\begin{eqnarray}
\mathcal{T}({\bf{v}})=\{i|v_{i}\ne0\},
\end{eqnarray}
where ${\bf{v}}=[v_i]$ is indexed by $i\in\mathcal{V}$ with $\mathcal{V}=\{1,\dots,
MN\}$. Furthermore, define the sets $\mathcal{\mathcal{V}}_{l}=\{M\sum_{i=1}^{l-1}N_{i}+1,\dots,M\sum_{i=1}^{l}N_{i}\},
l=1,\dots, L$, as a partition of $\mathcal{V}$, such that $\tilde{\bf{v}}_{l}=[v_{i}]$
is indexed by $i\in\mathcal{V}_{l}$.
The network power consumption in the first term of (\ref{npc}) thus
can be defined by the following \emph{combinatorial function} with respect
to the support of the beamforming vector, i.e.,
\begin{eqnarray}
F(\mathcal{T}({\bf{v}}))=\sum\limits_{l=1}^{L}P_{l}^c I(\mathcal{T}({\bf{v}})\cap\mathcal{V}_{l}\ne\emptyset),
\end{eqnarray}
where $I(\mathcal{T}\cap\mathcal{V}_{l}\ne\emptyset)$ is an indicator function
that takes value 1 if $\mathcal{T}\cap\mathcal{V}_{l}\ne\emptyset$ and 0
otherwise. Therefore, the total relative fronthaul link power consumption
can be reduced by encouraging the group-sparsity structure of the beamforming
vector $\bf{v}$. 

Furthermore, the total transmit power consumption in the second term
of (\ref{npc}) can be defined by the  \emph{continuous function} with respect
to the $\ell_2$-norms of the beamforming vector, i.e., 
\begin{eqnarray}
T({\bf{v}})=\sum_{l=1}^{L}\sum_{m=1}^{M}{1\over{\eta_{l}}}\|{\bf{v}}_{lm}\|_2^2,
\end{eqnarray} 
which implicates that the transmit powers of the inactive RRHs are zero, i.e., the corresponding beamforming coefficients are zero. 
Therefore, the transmit power consumption can be minimized by controlling
the magnitude of the beamforming coefficients. As a result,
the network power consumption in (\ref{npc}) can be rewritten as the following
combinatorial composite  function parameterized by the beamforming vector coefficients
${\bf{v}}$, i.e.,
\begin{eqnarray}
\label{comb}
P({\bf{v}})=F(\mathcal{T}({\bf{v}}))+T({\bf{v}}).
\end{eqnarray} 
Thus, it requires to \emph{simultaneously} control both the combinatorial  function $F$
and the continuous  function $T$ to minimize the network power consumption.
Such a composite function in (\ref{comb}) captures the unique property of
the network power consumption that involves two parts (i.e., relative fronthaul network power consumption and transmit power consumption) only through the beamforming coefficients ${\bf{v}}$.

\subsection{Group Sparse Beamforming Modeling}
Based on (\ref{comb}), problem $\mathscr{P}$  can be reformulated as the following robust group sparse beamforming  problem  
\begin{eqnarray} 
\label{rbf}
\mathscr{P}_{\textrm{sparse}}:
\mathop {\rm{minimize}}_{\bf{v}}&&F(\mathcal{T}({\bf{v}}))+T({\bf{v}})\nonumber\\
{\rm{subject~ to}} &&\sum_{m=1}^{M}\|{\bf{v}}_{lm}\|_2^2\le P_l, \forall
l\in\mathcal{L}\nonumber\\
&&{{|(\hat{\bf{h}}_{k}+{\bf{e}}_{k})^{\sf{H}}{\bf{v}}_{m}|^2}\over{\sum_{i\ne
m}|(\hat{\bf{h}}_{k}+{\bf{e}}_{k})^{\sf{H}}{\bf{v}}_{i}|^2+\sigma_{k}^2}}\ge
\gamma_k \nonumber \\
&&{\bf{e}}_{k}^{\sf{H}}{\bf{\Theta}}_{k}{\bf{e}}_{k}\le1,\forall k\in\mathcal{G}_m,
m\in\mathcal{M},
\end{eqnarray}
via optimizing the beamforming coefficients ${\bf{v}}$. We will show that the special structure of the objective function in $\mathscr{P}_{\textrm{sparse}}$ yields computationally efficient algorithm design. In particular, the weighted mixed $\ell_1/\ell_2$-norm will be derived as a convex surrogate to control both parts in (\ref{comb}) by inducing the group-sparsity structure for the robust multicast beamforming vector $\bf{v}$, thereby providing guidelines for RRH selection.

\section{A Semidefinite Programming Based Robust Group Sparse Beamforming Algorithm  }
\label{sdrfra}
\begin{figure}[t]
  \centering
  \includegraphics[width=1\columnwidth]{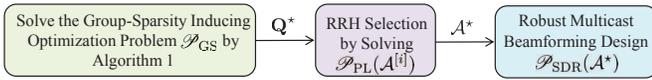}
 \caption{The proposed three-stage robust group sparse beamforming framework.}
  \label{gsbf}
 \end{figure}

In this section, we will present the semidefinite programming technique for the robust group sparse beamforming problem $\mathscr{P}_{\textrm{sparse}}$ by lifting the problem to higher dimensions. The general idea is to relax the combinatorial composite objective function  by the quadratic variational formulation of  the weighted mixed $\ell_1/\ell_2$-norm to induce the group-sparsity structure for the beamforming vector $\bf{v}$. Unfortunately, the resultant group sparse inducing optimization problem is still non-convex. We thus propose a perturbed alternating optimization algorithm to find a stationary point to it, thereby providing the information on determining the priority for the RRHs that should be switched off.  Based on the ordering result, a selection procedure
is then performed to determine active RRH sets, followed by the robust multicast coordinated
beamforming for the active RRHs in the final stage.  {\rev{The proposed three-stage robust group sparse beamforming framework is presented in Fig. {\ref{gsbf}}.}} 

\subsection{Stage One: 
Group-Sparsity Inducing Penalty  Minimization}
\label{gspa}
In this section, we describe a systematic way to address the combinatorial challenge in problem $\mathscr{P}_{\textrm{sparse}}$ by deriving a convex surrogate to approximate the composite objective function in problem $\mathscr{P}_{\textrm{sparse}}$. Specifically, we first derive the tightest convex positively homogeneous lower bound for the  network power consumption function (\ref{comb}) in the following proposition.

\begin{proposition} 
\label{gsdn}
The tightest convex positively homogeneous lower bound of  the objective
function in problem $\mathscr{P}_{\textrm{sparse}}$ is given by
\begin{eqnarray}
\label{plower}
\Omega({\bf{v}})=2\sum_{l=1}^{L}\sqrt{P_{l}^c\over{\eta_{l}}}\|\tilde{\bf{v}}_{l}\|_{2},
\end{eqnarray}
which is a group-sparsity inducing norm for the aggregative robust multicast beamformer vector ${\bf{v}}$.
\begin{IEEEproof}
Please refer to \cite[Appendix A]{Yuanming_TWC2014} for the proof.   
\end{IEEEproof}
\end{proposition}

Based on proposition {\ref{gsdn}}, we propose to minimize the  weighted mixed
$\ell_1/\ell_2$-norm to induce the group-sparsity structure for the aggregative robust multicast beamforming vector
${\bf{v}}$:
\begin{eqnarray}
\!\!\!\!\!\!\!\!\!\mathscr{P}_{\textrm{GSBF}}: 
\mathop {\rm{minimize}}_{\bf{v}}&&\Omega({\bf{v}})\nonumber\\
{\rm{subject~ to}} &&\sum_{m=1}^{M}\|{\bf{v}}_{lm}\|_2^2\le P_l, \forall
l\in\mathcal{L}\nonumber\\
\label{wcqos1}
&&{{|(\hat{\bf{h}}_{k}+{\bf{e}}_{k})^{\sf{H}}{\bf{v}}_{m}|^2}\over{\sum_{i\ne
m}|(\hat{\bf{h}}_{k}+{\bf{e}}_{k})^{\sf{H}}{\bf{v}}_{i}|^2+\sigma_{k}^2}}\ge
\gamma_k  \\
\label{wcqos2}
&&{\bf{e}}_{k}^{\sf{H}}{\bf{\Theta}}_{k}{\bf{e}}_{k}\le1,\forall k\in\mathcal{G}_m,
m\in\mathcal{M}.
\end{eqnarray}
This is, however, a non-convex optimization problem due to the non-convex
worst-case QoS constraints (\ref{wcqos1}) and (\ref{wcqos2}). 

To seek computationally efficient algorithms to solve the non-convex problem $\mathscr{P}_{\textrm{GSBF}}$,
we  propose to lift the problem to higher dimensions with optimization
variables as ${\bf{Q}}_{m}={\bf{v}}_{m}{\bf{v}}_{m}^{\sf{H}}\in\mathbb{C}^{N\times
N}, \forall m$. To achieve this goal, in Section {\ref{squ12}}, a variational
formulation is proposed  to turn the non-smooth group-sparsity
inducing norm $\Omega({\bf{v}})$ into a smooth one with quadratic forms, thereby extracting the variables
${\bf{Q}}_{m}$'s. We then
``linearize" the non-convex worst-case QoS constraints with the S-lemma in Section {\ref{sdrgs}}. In Section {\ref{alopgs}},
the perturbed alternating optimization algorithm is proposed
to solve the resultant non-convex group-sparsity inducing optimization
problem by exploiting its convex substructures.

\subsubsection{Quadratic Variational Formulation of the Weighted Mixed $\ell_1/\ell_2$-Norm}
\label{squ12}
In order to extract
the variables ${\bf{Q}}_{m}$'s from the weighted mixed $\ell_1/\ell_2$-norm, we introduce
the following lemma to obtain an equivalent expression
for the square norm $\Omega^{2}({\bf{v}})$, which has the
same capability of inducing group-sparsity as the non-smooth one $\Omega(\bf{v})$
\cite{Bach_ML2011} and is widely
used in multiple kernel learning \cite{Bach_ML2008}. 
\begin{lemma}
\label{qctr} 
\cite{Bach_ML2011} Let ${\bf{x}}=(x_1,\dots,x_L)\in\mathbb{R}_{+}^{L}$ and
${\bs{\omega}}=(\omega_1,\dots,\omega_L)\in\mathbb{R}_{+}^{L}$,
then
\begin{eqnarray}
\label{qct}
\left(\sum_{l=1}^{L}\omega_lx_{l}\right)^2=\inf_{{\boldsymbol{\mu}}\in\mathbb{R}_{+}^{L}}\sum_{l=1}^{L}{{\omega_l^2x_{l}^{2}}\over{\mu_{l}}},
~~{\rm{s. t.}} \sum_{l=1}^{L}\mu_{l}=1.
\end{eqnarray}
\end{lemma}
\begin{IEEEproof}
This can be obtained directly through the Cauchy-Schwarz inequality
\begin{eqnarray}
\sum_{l=1}^{L}\omega _lx_{l}&=&\sum_{l=1}^{L}{{\omega_lx_{l}}\over{\sqrt{\mu_{l}}}}\cdot\sqrt{\mu_{l}}\nonumber\\
&\le&\left(\sum_{l=1}^{L}{\omega_l^2x_{l}^2\over{\mu_{l}}}\right)^{1/2}\left(\sum_{l=1}^{L}\mu_{l}\right)^{1/2}\!\!\!\!\!,
\end{eqnarray}
where $\omega_l\ge0$ and the equality is met when $\sqrt{\mu_l}$ is proportional
to $(\omega_lx_l)/\sqrt{\mu_l}$, i.e.,
\begin{eqnarray}
\mu_l={{\omega _lx_l}\over{\sum_{l=1}^{L}\omega_lx_l}},
\end{eqnarray}
which leads to the conclusion (\ref{qct}). 
\end{IEEEproof}

Based on lemma {\ref{qctr}}, the square of the weighted mixed $\ell_1/\ell_2$-norm (\ref{plower})
can be rewritten as
\begin{eqnarray}
\Omega^{2}({\bf{v}})=\inf_{{\boldsymbol{\mu}}\in\mathcal{X}}\mathcal{R}(\boldsymbol{\mu},
{\bf{Q}}),
\end{eqnarray}
where $\mathcal{X}=\{\mu_{l}|~\mu_{l}>0,\sum_{l=1}^{L}{\mu_{l}}=1\}$
is a simplex set and
\begin{eqnarray}
\label{qnorm1}
\mathcal{R}(\boldsymbol{\mu}, {\bf{Q}})=4\sum_{l=1}^{L}{{P_{l}^c}\over{\eta_{l}\mu_{l}}}\left(\sum_{m=1}^{M}{\rm{Tr}}({\bf{C}}_{lm}{\bf{Q}}_{m})\right),
\end{eqnarray}
where ${\bs{\mu}}=[\mu_l]$, ${\bf{Q}}=[{\bf{Q}}_l]$ and ${\bf{C}}_{lm}\in\mathbb{R}^{N\times
N}$ is a block diagonal matrix
with the identity matrix ${\bf{I}}_{N_l}$ as the $l$-th main diagonal block
square matrix and zeros elsewhere. Therefore, the group-sparsity structure
of the beamformer ${\bf{v}}$ can be extracted from the trace of ${\bf{Q}}_m$'s,
as will be shown in (\ref{tracenorm}). This procedure is known as \emph{the
quadratic variational formulation} of norms \cite{Bach_ML2011}.

\subsubsection{Linearize the Non-convex Worst-case QoS Constraints}
\label{sdrgs}
Define ${\bf{G}}_{m}=({\bf{Q}}_{m}-\gamma_{k}\sum_{i\ne m}{\bf{Q}}_{i})$, and then the worst-case QoS constraints (\ref{wcqos1}) and (\ref{wcqos2}) can be rewritten as
\begin{eqnarray}
\label{wcqos}
\min_{{\bf{e}}_{k}^{\sf{H}}{\bf{\Theta}}_{k}{\bf{e}}_{k}\le1}(\hat{\bf{h}}_{k}+{\bf{e}}_{k})^{\sf{H}}{\bf{G}}_{m}(\hat{\bf{h}}_{k}+{\bf{e}}_{k})\ge\gamma_{k}\sigma_{k}^2,
\forall k\in\mathcal{G}_m. 
\end{eqnarray}
As the number of choices of ${\bf{e}}_{k}$'s in the worst-case QoS constraint
(\ref{wcqos}) is infinite, there are an infinite number of such ``linearized" QoS constraints.
Fortunately, using the S-lemma \cite[Appendix B.2]{boyd2004convex},
the worst-case QoS constraints (\ref{wcqos}) can be  equivalently written as the following
finite number of convex constraints:
\begin{eqnarray}
\!\!\!\!\!\!\!\mathcal{C}_1:
\left[ \begin{array}{c|c}
{\bf{G}}_{m} & {\bf{G}}_{m}\hat{{\bf{h}}}_{k} \\
\hline
{\hat{\bf{h}}}_{k}^{\sf{H}}{\bf{G}}_{m} & \hat{{\bf{h}}}_{k}^{\sf{H}}{\bf{G}}_{m}\hat{{\bf{h}}}_{k}-\gamma_{k}\sigma_{k}^2
 \\
\end{array} \right]+\lambda_{k}
\left[\begin{array}{c|c}
{\bf{\Theta}}_{k} & {\bf{0}}\\
\hline
{\bf{0}}^{\sf{H}} & -1
\end{array}\right]\succeq {\bf{0}}, 
\end{eqnarray}
where $\lambda_{k}\ge 0$ and $ k\in\mathcal{G}_m$ with $m\in\mathcal{M}$.

Based on the above discussions and utilizing the principle of SDR technique \cite{Z.Q.Luo_SPM2010} by dropping the rank-one constraints for ${\bf{Q}}_k$'s,  we propose to solve the following problem  to induce the group-sparsity structure for the beamforming vector $\bf{v}$ 
\begin{eqnarray}\label{gsbf1}
\mathscr{P}_{\textrm{GS}}: 
\mathop {\textrm{minimize}}_{{\bf{Q}}, {\boldsymbol{\lambda}}, {\boldsymbol{\mu}}\in\mathcal{X}}&&
\mathcal{R}(\boldsymbol{\mu}, {\bf{Q}}) \nonumber\\
\textrm{subject to}&& \mathcal{C}_{1}, \mathcal{C}_2(\mathcal{L}), \lambda_{k}\ge0,
{{\bf{Q}}_{m}\succeq{\bf{0}}}\nonumber\\
&&\forall k\in\mathcal{G}_m, m\in\mathcal{M},
\end{eqnarray}
where ${\bs{\lambda}}=[\lambda_k]$ and $\mathcal{C}_2(\mathcal{A})$ is the set of linearized per-RRH transmit power constraints,
\begin{eqnarray}
\mathcal{C}_2(\mathcal{A}): \sum_{m=1}^{M}{\rm{Tr}}({\bf{C}}_{lm}{\bf{Q}}_{m})\le P_l, l\in\mathcal{A}.
\end{eqnarray}
Problem $\mathscr{P}_{\textrm{GS}}$ is still non-convex, as the objective function $\mathcal{R}({\bs{\mu}}, {\bf{Q}})$ is
not jointly convex in the variables (${\bs{\mu}}, {\bf{Q}}$). Nevertheless, the objective function is biconvex \cite{Gorski2007biconvex}, i.e., function $\mathcal{R}$ is convex with respect to $\bs{\mu}$ for fixed ${\bf{Q}}$ and vice versa.  In the next subsection, we thus exploit the convex substructures of problem $\mathscr{P}_{\textrm{GS}}$ to develop a perturbed alternating optimization algorithm to find an efficient sub-optimal solution.

\subsubsection{Perturbed Alternating Optimization Algorithm}
\label{alopgs}
The general idea of the alternating optimization algorithm is that problem $\mathscr{P}_{\textrm{GS}}$ is first optimized with respect to $({\bf{Q}},{\boldsymbol{\lambda}})$
with a fixed  ${\boldsymbol{\mu}}$, then the variables ${{\mu}}_l$'s are chosen to minimize ${\mathcal{R}}({\boldsymbol{\mu}}, {\bf{Q}})$ with a fixed ${\bf{Q}}$. However, to avoid singularity when $\mu_l$'s approach to zeros during the alternating procedure as discussed in \cite{Bach_ML2008},
we instead adopt the  perturbed version of the alternating optimization
algorithm \cite{Argyriou_2008convexmutitask} to solve problem $\mathscr{P}_{\textrm{GS}}$.
Specifically, define the perturbed objective function of problem $\mathscr{P}_{\textrm{GS}}$
as
\begin{eqnarray}
\mathcal{R}_{\epsilon}(\boldsymbol{\mu}, {\bf{Q}})=4\sum_{l=1}^{L}{{P_{l}^c}\over{\eta_{l}\mu_{l}}}\left(\sum_{m=1}^{M}{\rm{Tr}}({\bf{C}}_{lm}{\bf{Q}}_{m}+{\epsilon}{\bf{I}}_N)\right),
\end{eqnarray}
where $\epsilon>0$. Let $\mathscr{P}_{\textrm{GS}}(\epsilon)$ be the problem by replacing the objective function in problem $\mathscr{P}_{\textrm{GS}}$ with the perturbed function $\mathcal{R}_{\epsilon}(\boldsymbol{\mu}, {\bf{Q}})$. We thus solve problem $\mathscr{P}_{\textrm{GS}}(\epsilon)$ via alternatively solving the following two problems:
\begin{itemize}
\item {\emph{Fixing $\bs{\mu}$, Optimizing  ${\bf{Q}}$ and ${\bs{\lambda}}$}}:
Given ${\bs{\mu}}={\bs{\mu}}^{[i]}$ at the $i$-th iteration, we need to solve
the following problem
\begin{eqnarray}\label{gsbf12}
\mathscr{P}_{\textrm{GS}}^{[i]}(\epsilon; {\bs{\mu}}^{[i]}): 
\mathop {\textrm{minimize}}_{{\bf{Q}}, {\boldsymbol{\lambda}}}&&
\mathcal{R}_{\epsilon}(\boldsymbol{\mu}^{[i]}, {\bf{Q}}) \nonumber\\
\textrm{subject to}&& \mathcal{C}_{1}, \mathcal{C}_2(\mathcal{L}), \lambda_{k}\ge0,
{{\bf{Q}}_{m}\succeq{\bf{0}}}\nonumber\\
&&\forall k\in\mathcal{G}_m, m\in\mathcal{M},
\end{eqnarray}
to obtain $({\bf{Q}}^{[i]}, {\bs{\lambda}}^{[i]})$. This is an SDP problem and can be solved efficiently using the interior-point method \cite{boyd2004convex}. 
\item {\emph{Fixing ${\bf{Q}}$ and ${\bs{\lambda}}$, Optimizing  $\bs{\mu}$}}:
Given ${\bf{Q}}={\bf{Q}}^{[i]}$
at the $i$-th iteration, we need to optimize ${\bs{\mu}}$ over the simplex set $\mathcal{X}$, i.e.,
\begin{eqnarray}
\mathscr{P}_{\textrm{GS}}^{[i]}(\epsilon; {\bf{Q}}^{[i]}): 
\mathop {\textrm{minimize}}_{{\bs{\mu}}\in\mathcal{X}}&&
\mathcal{R}_{\epsilon}(\boldsymbol{\mu}, {\bf{Q}}^{[i]}),
\end{eqnarray}
which has the following optimal solution based on Lemma \ref{qctr}:
\begin{eqnarray}
\label{simplexup}
\!\!\!\!\!\!\!\!\!\!\!\!\!{{\mu}}_{l}^{[i]}={\sqrt{(P_l^c/\eta_l)\cdot\sum_{m=1}^{M}{\rm{Tr}}({\bf{C}}_{lm}{\bf{Q}}_m^{[i]}+{\epsilon}{\bf{I}}_N)}\over{\sum_{l=1}^{L}\sqrt{(P_l^c/\eta_l)\cdot\sum_{m=1}^{M}{\rm{Tr}}({\bf{C}}_{lm}{\bf{Q}}_{m}^{[i]}+\epsilon{\bf{I}}_N)}}},
\end{eqnarray}
for any $l\in\mathcal{L}$.
\end{itemize}

As the objective function in problem $\mathscr{P}_{\textrm{GS}}(\epsilon)$ is bounded and non-increasing at each iteration,  the sequence $\{\mathcal{R}_{\epsilon}({\bs{\mu}}^{[i]}, {\bf{Q}}^{[i]})\}$ generated by this algorithm, clearly, converges monotonically to a sub-optimal value \cite{Gorski2007biconvex}. Since we will use the solution of the problem $\mathscr{P}_{\textrm{GS}}(\epsilon)$ to predicate the group-sparsity pattern for the beamformer $\bf{v}$, we thus are also interested in investigating the convergence of the sequence $\{{\bf{Q}}^{[i]}\}$ itself generated by this algorithm when $\epsilon\rightarrow 0$. This is presented in the following Theorem.
\begin{theorem}
\label{conal}
The sequence $\{{\bs{\mu}}^{[i]}(\epsilon), {\bf{Q}}^{[i]}(\epsilon), {\bs{\lambda}}^{[i]}(\epsilon)\}$ generated by the perturbed alternating optimization algorithm converges to a stationary point of problem $\mathscr{P}_{\textrm{GS}}(\epsilon)$.
Furthermore, when $\epsilon\rightarrow 0$, we have 
\begin{eqnarray}
\label{epscon}
\lim_{\epsilon\searrow\ 0}\mathbb{D}(\Lambda(\epsilon), \Lambda_{0})=0,
\end{eqnarray}
where $\Lambda_{0}$ (\ref{sta1}) and $\Lambda(\epsilon)$ (\ref{sta2})  denote the set of stationary points of problem $\mathscr{P}_{\textrm{GS}}$ and $\mathscr{P}_{\textrm{GS}}(\epsilon)$, respectively; and $\mathbb{D}(A_1,A_2)$, defined in (\ref{dist}), denotes the deviation of the set $A_1$ from the set $A_2$.
\begin{IEEEproof}
Please refer to Appendix {\ref{conprof}} for details.
\end{IEEEproof} 
\end{theorem}

The perturbed alternating optimization algorithm is presented in Algorithm {\ref{pal}}.
\begin{algorithm}
\label{pal}
\caption{Perturbed Alternating Optimization Algorithm }
{\textbf{input}}: Initialize ${\bs{\mu}}^{[0]}=(1/L, \dots, 1/L)$;
$I$ (the maximum number of iterations)\\
 Repeat
\begin{enumerate}
\item Solve problem $\mathscr{P}_{\textrm{GS}}^{[i]}(\epsilon; {\bs{\mu}}^{[i]})$ (\ref{gsbf12}). If it is feasible,\\
{\textbf{go to}} 2); otherwise, {\textbf{stop}} and return {\textbf{output 2}}.
\item Calculate ${\bs{\mu}}^{[i]}=(\mu_1^{[i]},\dots, {\mu}_{L}^{[i]})$
according to (\ref{simplexup}).
\end{enumerate}
 Until convergence or attain the  maximum iterations and return {\textbf{output 1}}.\\
 {\textbf{output 1}}: ${\bf{Q}}_1^{\star}, \dots, {\bf{Q}}_M^{\star}$; {\textbf{output 2}}: Infeasible.
\label{gsbfal} 
\end{algorithm} 

Based on the solutions ${\bf{Q}}_m^{\star}$'s generated by the perturbed alternating optimization algorithm, in the next subsection,
we will present how to extract the group-sparsity pattern information for the beamformer ${\bf{v}}$, thereby providing information on  the RRH ordering, i.e., determine the priority of the RRHs that should be switched
off.

\subsection{Stage Two: RRH Selection}
\label{rrhsel}

Given the solution ${{\bf{Q}}}^{\star}$
to the group sparse inducing optimization problem $\mathscr{P}_{\textrm{GS}}$,
 the group-sparsity structure information for the beamformer ${\bf{v}}$ can
be extracted from the following relation:
\begin{eqnarray}
\label{tracenorm}
\|\tilde{\bf{v}}_l\|_{\ell_2}=\left(\sum_{m=1}^{M}{\rm{Tr}}({\bf{C}}_{lm}{\bf{Q}}_{m})\right)^{1/2},
\forall l.
\end{eqnarray} 
Based on the (approximated) group-sparsity information in (\ref{tracenorm}),
the following ordering criterion \cite{Yuanming_TWC2014} incorporating
the key system parameters is adopted to determine
which RRHs should
be switched off, i.e.,
\begin{eqnarray}
\label{sparse_selection1}
\theta_{l}=\sqrt{{\kappa_{l}\eta_{l}}\over{P_{l}^c}}\left(\sum_{m=1}^{M}{\rm{Tr}}({\bf{C}}_{lm}{\bf{Q}}_{m}^{\star}\right)^{1/2},
\forall l\in\mathcal{L},
\end{eqnarray}
where $\kappa_{l}=\sum\nolimits_{k=1}^{K}\|\hat{\bf{h}}_{kl}\|_{2}^2$
is the channel gain for the estimated channel coefficients between
RRH $l$ and all the MUs. Therefore, the RRH with a smaller parameter $\theta_{l}$
will have a higher priority to be switched off. Note that most previous works
applying
the idea of sparsity inducing norm minimization approach directly map the
sparsity pattern to their applications. For instance, in \cite{Mehanna_SP2013},
the transmit
antenna with smaller coefficients in the beamforming coefficient group (measured
by the $\ell_{\infty}$-norm) will have a higher priority to be switched off.
In \cite{Yuanming_TWC2014}, however, we show that the ordering rule ({\ref{sparse_selection1}}), which incorporates the key system parameters, 
yields much better performance than the pure sparsity pattern based selection
rule in terms of network power minimization. 

In this paper, we adopt a simple RRH selection procedure, i.e.,  binary search,
due to its low-complexity. Specifically,
based on the ordering rule (\ref{sparse_selection1}), we sort the coefficients
 in the ascending order: $\theta_{\pi_{1}}\le\theta_{\pi_{2}}\le\cdots\le\theta_{\pi_{L}}$
to determine the active RRH set. {\rev{Denote $J_0$ as the maximum number of RRHs that can be switched off. That is, problem
$\mathscr{F}(\mathcal{A}^{[i]})$
is feasible for any $i\le J_{0}$, }}
\begin{eqnarray}
\label{feasibility}
\mathscr{F}(\mathcal{A}^{[i]}):
\mathop {\textrm{find}}&&{\bf{v}}\nonumber\\
{\textrm{subject to}}&&(\ref{robustc1}), (\ref{pz}), (\ref{qos1}), (\ref{qos2}),
\end{eqnarray}
 where $\mathcal{A}^{[i]}\cup\mathcal{Z}^{[i]}=\mathcal{L}$
with $\mathcal{Z}^{[i]}=\{{\pi_{0}},{\pi_{1}},\dots,{\pi_{i}}\}$ and $\pi_{0}=\emptyset$. {\rev{Likewise, problem $\mathscr{F}(\mathcal{A}^{[i]})$ with $\mathcal{A}^{[i]}=\{\pi_{i+1},\dots, \pi_{L}\}$ is infeasible for any $i> J_0$.}} A binary search procedure can be adopted to determine $J_{0}$, which only
needs to solve no more than $(1+\lceil\log(1+L)\rceil)$ feasibility problems (\ref{feasibility}) as will be presented in Algorithm {\ref{rgsbff}}. Denote $\mathcal{A}^{[J_{0}]}$ as the final active RRH
set, we thus need to solve the following transmit power minimization problem
 \begin{eqnarray}
\mathscr{P}(\mathcal{A}):
\mathop {\textrm{minimize}}_{\bf{v}}&&\sum\limits_{l\in\mathcal{A}}\left({1\over{\eta_{l}}}\sum_{m=1}^{M}\|{\bf{v}}_{lm}\|_2^2+P_{l}^{c}\right)\nonumber\\
\textrm{subject to}&&(\ref{robustc1}), (\ref{pz}), (\ref{qos1}), (\ref{qos2}),
\end{eqnarray} 
with the fixed active RRH set $\mathcal{A}=\mathcal{A}^{[J_0]}$ to determine
the  transmit beamformer coefficients for the active RRHs. Unfortunately, both problems $\mathscr{F}(\mathcal{A})$
and $\mathscr{P}(\mathcal{A})$ are non-convex and intractable. Thus, in the
paper, we resort to the computationally efficient semidefinite programming technique to find approximate
solutions to feasibility problem $\mathscr{F}(\mathcal{A})$ and optimization
problem $\mathscr{P}(\mathcal{A})$. 

Notice that, with perfect CSI assumptions as in \cite{Yuanming_TWC2014, Mehanna_SP2013}, given the active RRH set $\mathcal{A}$, the size of the corresponding
optimization problem $\mathscr{P}(\mathcal{A})$ (e.g., \cite[(12)]{Yuanming_TWC2014} and \cite[(13)]{Mehanna_SP2013}) will be reduced. The key observation is that we only need to consider the
channel links from the active RRHs. However, with imperfect CSI, we still
need to consider the channel links from all the RRHs due to the lack of the knowledge of the exact values of the CSI errors ${\bf{e}}_k$'s. As a result, the sizes of corresponding optimization problems $\mathscr{P}(\mathcal{A}^{[i]})$'s cannot be reduced with imperfect CSI. 

\subsubsection{PhaseLift to the Non-convex Feasibility Problem}
\label{phaselift}
In this subsection, we use the \emph{PhaseLift} technique \cite{candes_2012phaseretrieval} to find approximate solutions to the
non-convex feasibility problem $\mathscr{F}(\mathcal{A})$. Specifically,
we first lift
the problem to higher dimensions such that  the feasibility problem $\mathscr{F}(\mathcal{A})$
can be reformulated as
\begin{eqnarray}
\label{liftfeasible}
\mathop {\textrm{find}}&& {\bf{Q}}_1,\dots, {\bf{Q}}_M\nonumber\\
\textrm{subject to}&&\mathcal{C}_1, \mathcal{C}_2(\mathcal{A}), \mathcal{C}_3(\mathcal{Z}),
\lambda_{k}\ge0, {\bf{Q}}_{m}\succeq{\bf{0}}\nonumber\\
\label{sdprankone}
&&~{\rm{rank}}({\bf{Q}}_{m})=1,
\forall k\in\mathcal{G}_m, m\in\mathcal{M},
\end{eqnarray}
where
\begin{eqnarray}
\mathcal{C}_3(\mathcal{Z}):
\sum\limits_{m=1}^{M}{\rm{Tr}}({\bf{C}}_{lm}{\bf{Q}}_{m})=0,
\forall l\in\mathcal{Z}.
\end{eqnarray}
The main idea of the PhaseLift technique is to approximate the non-convex rank functions in problem (\ref{liftfeasible}) using the convex surrogates, yielding the following convex feasibility problem
\begin{eqnarray}
\label{phlift1}
\mathscr{P}_{\textrm{PL}}(\mathcal{A}):
\mathop {\textrm{find}}&& {\bf{Q}}_1,\dots, {\bf{Q}}_M\nonumber\\
\textrm{subject to}&&\mathcal{C}_1, \mathcal{C}_2(\mathcal{A}), \mathcal{C}_3(\mathcal{Z}),
\lambda_{k}\ge0, {\bf{Q}}_{m}\succeq{\bf{0}}\nonumber\\
&&~\forall k\in\mathcal{G}_m, m\in\mathcal{M},
\end{eqnarray}
which is an SDP problem and can be solved using
the interior-point method \cite{boyd2004convex} efficiently. In general, the solution of problem  $\mathscr{P}_{\textrm{PL}}(\mathcal{A})$ may not be rank-one. If this happens, to yield a feasible solution for problem $\mathscr{F}(\mathcal{A})$, the Gaussian randomization procedure \cite{Z.Q.Luo_SPM2010} will be applied to obtain a feasible rank-one approximate solution for problem $\mathscr{F}(\mathcal{A})$ from the solution of problem $\mathscr{P}_{\textrm{PL}}(\mathcal{A})$.

\begin{remark}
The PhaseLift technique, {\rev{serving as one promising application of the SDR method}}, was proposed in \cite{candes_2012phaseretrieval} to solve
the phase retrieval problem \cite{Candes2013phase}, which is mathematically
a feasibility problem with multiple quadratic equation constraints. Various
conditions are presented in \cite{candes_2012phaseretrieval, Candes2013phase} for {\rev{the phase retrieval problem}}, under which
the corresponding solution of the PhaseLift relaxation problem yields a rank-one
solution with a high probability. However, for our problem $\mathscr{P}_{\textrm{PL}}$
with additional complicated constraints, it is challenging to perform such rank-one
solution analysis. Thus, in this paper, we only focus on developing computationally efficient approximation
algorithms {\rev{based on the SDR technique}}.
\end{remark}

\subsection{Stage Three: SDR to the Robust Multicast Beamforming Problem}
\label{sdr}
Once we have selected active RRHs, i.e., fix the set $\mathcal{A}$, we need to finalize the beamforming vector by solving problem $\mathscr{P}(\mathcal{A})$. We lift the non-convex optimization problem $\mathscr{P}(\mathcal{A})$
to higher dimensions and adopt the SDR
technique  by dropping the rank-one constraints, yielding
the following convex relaxation problem 
\begin{eqnarray}
\mathscr{P}_{\textrm{SDR}}(\mathcal{A}):
\mathop {\textrm{minimize}}_{{\bf{Q}}, {\boldsymbol{\lambda}}}&&~
\sum\limits_{l\in\mathcal{A}}\left({1\over{\eta_{l}}}\sum\limits_{m=1}^{M}{\rm{Tr}}({\bf{C}}_{lm}{\bf{Q}}_{m})+P_{l}^{c}\right)\nonumber\\
\textrm{subject to}&&~\mathcal{C}_1, \mathcal{C}_2(\mathcal{A}), \mathcal{C}_3(\mathcal{Z}),
\lambda_{k}\ge0, {\bf{Q}}_{m}\succeq{\bf{0}}\nonumber\\
&&~\forall k\in\mathcal{G}_m, m\in\mathcal{M},
\label{SDPP2}
\end{eqnarray}
which is an SDP problem and can be solved using the interior-point method \cite{boyd2004convex}. It is important to investigate whether the solution
of  problem $\mathscr{P}_{\textrm{SDR}}(\mathcal{A})$ yields a rank-one
solution $\{{\bf{Q}}_m^{\star}\}$. This is, however, an on-going research topic and some preliminary results were presented in \cite{shen2012distributed,Z.Q.Luo_2012robust}. In this paper, if  ${\rm{rank}}({\bf{Q}}_m^{\star})=1, \forall m$, we can write ${\bf{Q}}_m^{\star}={\bf{v}}_m^{\star}{\bf{v}}_m^{\star
\sf{H}}, \forall m$ and $\{{\bf{v}}_m^{\star}\}$ is a feasible (in fact optimal)
solution to problem $\mathscr{P}(\mathcal{A})$. Otherwise, if the rank-one
solution is failed to be obtained, the Gaussian randomization method \cite{Z.Q.Luo_SPM2010}
will be employed to obtain a feasible rank-one  approximate solution to problem
$\mathscr{P}(\mathcal{A})$. 

Finally, we arrive at the robust group sparse beamforming algorithm
as shown in Algorithm {\ref{rgsbff}}.
 \begin{algorithm}
\label{rgsbff}
\caption{Robust Group Sparse Beamforming Algorithm}
\textbf{Step 0:} Solve the group-sparsity inducing  optimization
problem $\mathscr{P}_{\textrm{GS}}$ (\ref{gsbf1}) using Algorithm 1.\\
\begin{enumerate}
\item {\textbf{If}} it is infeasible, {\textbf{go
to End}}.
\item  {\textbf{If}} it is feasible, obtain the solutions ${\bf{Q}}_m^{\star}$'s,
calculate the ordering criterion  (\ref{sparse_selection1}), and sort them in
the ascending order: $\theta_{\pi_{1}}\le\dots\le\theta_{\pi_{L}}$, {\textbf{go
to Step 1}}.
\end{enumerate}
\textbf{Step 1:} Initialize $J_{\textrm{low}}=0$, $J_{\textrm{up}}=L$, $i=0$.\\
\textbf{Step 2:} Repeat
\begin{enumerate}
\item Set $i\leftarrow\lfloor{{J_{\textrm{low}}+J_{\textrm{up}}}\over{2}}\rfloor$.\\
\item Solve  problem $\mathscr{P}_{\textrm{PL}}(\mathcal{A}^{[i]})$ (\ref{phlift1}):
if it is infeasible, set $ J_{\textrm{up}}=i$; otherwise, set $J_{\textrm{low}}=i$.
\end{enumerate}
\textbf{Step 3:} Until $J_{\textrm{up}}-J_{\textrm{low}}=1$, obtain $J_{0}=J_{\textrm{low}}$
and obtain the optimal active RRH set $\mathcal{A}^{\star}$ with $\mathcal{A}^{\star}\cup\mathcal{J}=\mathcal{L}$
and $\mathcal{J}=\{{\pi_{1}},\dots, {\pi_{J_{0}}}\}$.\\
\textbf{Step 4:} Solve problem $\mathscr{P}_{\textrm{SDR}}(\mathcal{A}^{\star})$ (\ref{SDPP2}), obtain the robust multicast beamforming coefficients for the active RRHs.\\
\textbf{End}
\end{algorithm}

{\rev{\begin{remark}
The proposed robust group sparse beamforming algorithm consists of
three stages. In the first stage, we observe that the perturbed alternating
optimization algorithm converges
in 20 iterations on average in all the simulated settings in this paper, while it is interesting to analyze the convergence rate for this algorithm.
In the second stage, to find the set of active RRHs, we only need to solve
no more than $(1+\lceil\log(1+L)\rceil)$ convex feasibility problems (\ref{phlift1}) using the bi-section method.
Finally, we need to solve problem (\ref{SDPP2}) to determine the transmit
beamforming coefficients for the fixed active RRHs.
\end{remark}}}

\section{Simulation Results}
\label{simu}

In this section, we analyze the performance of the  proposed robust group sparse beamforming algorithm.   For illustration purposes, all the estimated channels $\hat{\bf{h}}_k$'s are modeled
as spatially uncorrelated Rayleigh fading and the CSI errors are
modeled as the elliptic model (\ref{ellp}) with ${\bf{Q}}_{k}={\varepsilon}_k^{-2}{\bf{I}}_{N},
\forall k$.
We assume that each multicast group has the same number of MUs, i.e., $|\Omega_1|=|\Omega_2|=\dots=|\Omega_M|$. The power amplifier efficiency coefficients are set to be $\eta_l=25\%, \forall l$. The perturbed parameter $\epsilon$ in the perturbed alternating optimization algorithm is set to be $10^{-3}$ and the algorithm will stop if either the difference
between the objective values of consecutive iterations is less than $10^{-3}$ or it exceeds the predefined maximum iterations $20$.  Each point of the simulation results is averaged over 50 randomly generated channel realizations, except for Fig. {\ref{convergence}}, where we only report one typical channel realization.

\subsection{Convergence of the Perturbed Alternating Optimization Algorithm}
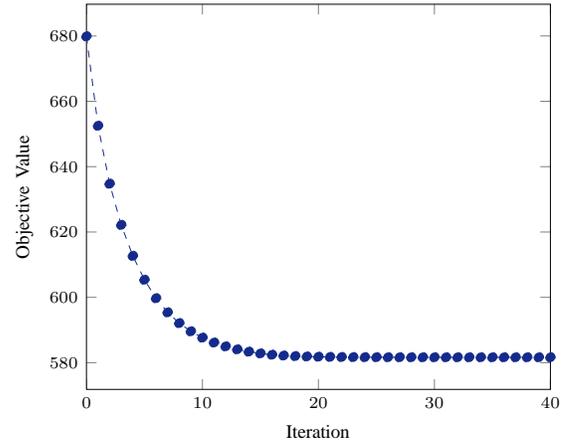
\begin{figure}[!t]
\centering
\begin{tikzpicture}[scale=0.9]
\begin{axis}[
xmin=0, xmax=40, 
xlabel={Iteration}, 
ylabel={Objective Value}, 
every axis x label/.style={at={(ticklabel cs:0.5)},anchor=near
ticklabel},
every axis y label/.style={at={(ticklabel cs:0.5)},rotate=90, anchor=near
ticklabel}, 
label style={font=\footnotesize},
tick label style={font=\scriptsize},
legend style={at={(0,1)}, anchor=north west, font=\scriptsize}]
\addplot[colorhkust, style=dashed, mark=*] table {Convergence.dat};
\end{axis}
\end{tikzpicture}
\caption{Convergence of the perturbed alternating optimization algorithm.}
\label{convergence}
\end{figure}
Consider a network with $L=10$ 2-antennas RRHs and 3 multicast groups with
2 single-antenna MUs in each group, i.e., $|\Omega_m|=2, \forall m$. All error radii $\varepsilon_k$'s are set to be  $0.05$. The convergence
of the  perturbed alternating optimization algorithm is demonstrated in Fig. {\ref{convergence}} for a typical channel realization. This figure shows that the proposed alternating optimization algorithm converges very fast (less 20 iterations) in the simulated network size.

\subsection{Network Power Minimization}
\subsubsection{Scenario One}
\begin{figure}[!t]
\centering
\begin{tikzpicture}[scale=0.9]
\begin{axis}[
xmin=0, xmax=8, 
xlabel={Target SINR [dB]}, 
ylabel={Average Network Power Consumption [W]}, 
every axis x label/.style={at={(ticklabel cs:0.5)},anchor=near
ticklabel},
every axis y label/.style={at={(ticklabel cs:0.5)},rotate=90, anchor=near
ticklabel}, 
label style={font=\footnotesize},
tick label style={font=\scriptsize},
legend style={at={(0,1)}, anchor=north west, font=\scriptsize}]
\addplot[purple, mark=diamond,  line width=1pt] table {TotalPower_CBF_data.dat};
\addplot[colorhkust, mark=square, mark size=1.4pt,line width=1pt] table
{TotalPower_Baseline_data.dat};
\addplot[violet, mark=o, mark size=1.8pt,line width=1pt] table {TotalPower_GSBFAlternating_data.dat};
\addplot[color1, mark=x,  line width=1pt] table {TotalPower_Exhaustive_data.dat};

\legend{[right]Coordinated Beamforming \cite{WeiYu_WC10},[right]$\ell_1/\ell_\infty$-Norm
 Algorithm \cite{Mehanna_SP2013}, [right]Proposed Algorithm, [right]Exhaustive
Search}
\end{axis}
\end{tikzpicture}
\caption{Average network power consumption versus target SINR for scenario one.}
\label{small}
\end{figure}
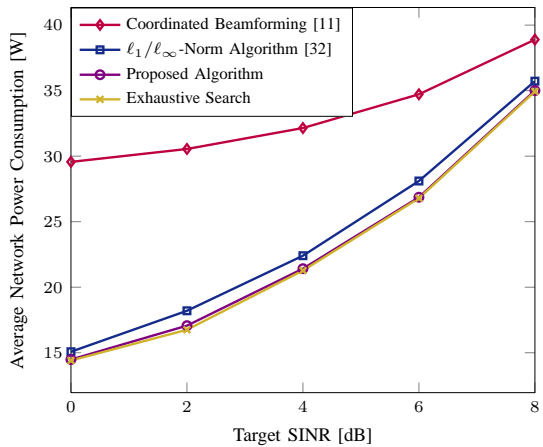
\begin{table}[!t]
\renewcommand{\arraystretch}{1.3}
\caption{The Average Number of Active RRHs with Different Algorithms for
Scenario One}
\label{sparsity1}
\centering
\begin{tabular}{l|c|c|c|c|c}
\hline
\tabincell{c}{{Target SINR [dB]}} & 0 & 2 &
4 & 6 & 8 \\
\hline
Coordinated Beamforming & 5.00 & 5.00 & 5.00 & 5.00 & 5.00  \\
\hline
\tabincell{c}{$\ell_1/\ell_\infty$-Norm  Algorithm} & 2.00 & 2.33 & 2.73
&
3.30 & 4.10 \\
\hline
\tabincell{c}{Proposed Algorithm} & 2.00 & 2.13 & 2.63 &
3.13 & 4.00  \\
\hline
\tabincell{c}{Exhaustive Search} & 2.00 & 2.07 & 2.60 & 3.10 & 4.00\\
\hline
\end{tabular}
\label{rrh_small}
\end{table} 

\begin{table}[!t]
\renewcommand{\arraystretch}{1.3}
\caption{The Average Total Transmit Power Consumption with Different Algorithms
for Scenario One}
\label{sparsity1}
\centering
\begin{tabular}{l|c|c|c|c|c}
\hline
\tabincell{c}{{Target SINR [dB]}} & 0 & 2 &
4 & 6 & 8 \\
\hline
Coordinated Beamforming & 1.56 & 2.55 & 4.15 & 6.72 & 10.89  \\
\hline
\tabincell{c}{$\ell_1/\ell_\infty$-Norm  Algorithm} & 3.88 & 5.13 & 7.10
&
9.63 & 12.76 \\
\hline
\tabincell{c}{Proposed Algorithm} & 3.28 & 5.12 & 6.67 &
9.32 & 12.61  \\
\hline
\tabincell{c}{Exhaustive Search} & 3.20 & 5.18 & 6.71 & 9.43 & 12.54\\
\hline
\end{tabular}
\label{transmit_small}
\end{table}

We first consider a network with $L=5$ 2-antenna RRHs and $M=2$ multicast groups each has 2 single-antenna MUs, i.e., $|\Omega_m|=2, \forall m$. The relative fronthaul links power consumption are set to be $P_{l}^{c}=5.6W, \forall l$. All error radii $\varepsilon_k$'s are set to be  $0.01$. Fig. {\ref{small}} demonstrates  the average network power consumption with different target SINRs. The corresponding
average number of  active RRHs and average total transmit power consumption are showed in Table {\ref{rrh_small}} and Table {\ref{transmit_small}}, respectively. 

Specifically, Fig. {\ref{small}} shows that the proposed robust group sparse beamforming algorithm achieves near-optimal values of network power consumption compared with the ones obtained by the exhaustive search algorithm {\rev{via solving a sequence of problems (\ref{SDPP2})}}. Furthermore, it is observed that the proposed algorithm outperforms the square of $\ell_1/\ell_\infty$-norm based algorithm with sparsity pattern ordering rule in \cite{Mehanna_SP2013} in terms of network power minimization. Specifically, the objective function of the group-sparsity inducing optimization problem (\ref{gsbf1}) will be replaced by $\mathcal{R}=\sum_{l_1=1}^{L}\sum_{l_2=1}^{L}\max_{m}\max_{n_{l_1}}\max_{n_{l_2}}|{\bf{Q}}_m(n_{l_1},n_{l_2})|$ with ${\bf{Q}}_{m}(i,j)$ being the entry indexed by $(i,j)$ in ${\bf{Q}}_m$. Then the RRH with smaller beamforming coefficients measured by the $\ell_\infty$-norm  will have a higher priority to be switched off. In particular, Table {\ref{rrh_small}} shows that the proposed algorithm can switch off  more RRHs than the $\ell_1/\ell_\infty$-norm based algorithm, which is almost the same as the exhaustive search algorithm. Besides, this table also verifies the group-sparsity assumption for the aggregative transmit beamformer $\bf{v}$, i.e., the beamforming coefficients of the switched off RRHs are set to be zeros simultaneously.  Meanwhile, Table {\ref{transmit_small}} shows that the proposed algorithm can achieve higher transmit beamforming gains, yielding lower  total transmit power consumption compared with the $\ell_1/\ell_\infty$-norm based algorithm. The coordinated beamforming algorithm \cite{WeiYu_WC10}, which aims at only minimizing the total transmit power consumption with all the RRHs active, achieves the highest beamforming gain but with the highest relative fronthaul links power consumption. 

Overall, Fig. {\ref{small}}, Table {\ref{rrh_small}} and Table {\ref{transmit_small}} show the effectiveness of the proposed robust group sparse beamforming algorithm to minimize the network power consumption.

\subsubsection{Scenario Two} 
\begin{figure}[t]
\centering
\begin{tikzpicture}[scale=0.9]
\begin{axis}[
xmin=0, xmax=8, 
xlabel={Target SINR [dB]}, 
ylabel={Average Network Power Consumption [W]}, 
every axis x label/.style={at={(ticklabel cs:0.5)},anchor=near
ticklabel},
every axis y label/.style={at={(ticklabel cs:0.5)},rotate=90, anchor=near
ticklabel}, 
label style={font=\footnotesize},
tick label style={font=\scriptsize},
legend style={at={(0,1)}, anchor=north west, font=\scriptsize}]
\addplot[purple, mark=diamond,  line width=1pt] table {TotalPower_CBF_large.dat};
\addplot[colorhkust, mark=square, mark size=1.4pt,line width=1pt] table {TotalPower_Baseline_large.dat};
\addplot[violet, mark=o, mark size=1.8pt,line width=1pt] table {TotalPower_GSBFAlternating_large.dat};

\legend{[right]Coordinated Beamforming \cite{WeiYu_WC10},[right] $\ell_1/\ell_\infty$-Norm
Algorithm \cite{Mehanna_SP2013}, [right]Proposed Algorithm}
\end{axis}
\end{tikzpicture}
\caption{Average network power consumption versus target SINR for scenario two.}
\label{large}
\end{figure}
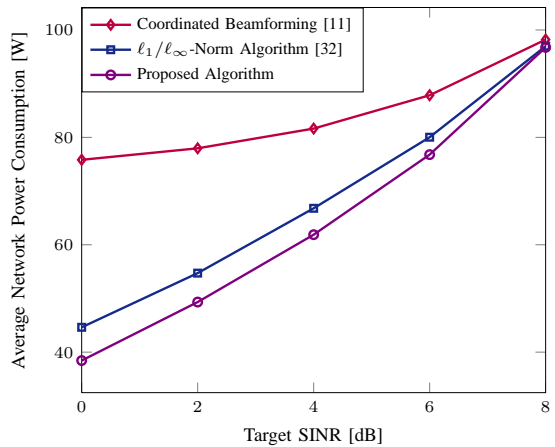

\begin{table}[!t]
\renewcommand{\arraystretch}{1.3}
\caption{The Average Relative Fronthaul Links Power Consumption with Different Algorithms for
Scenario Two}
\label{sparsity1}
\centering
\begin{tabular}{l|c|c|c|c|c}
\hline
\tabincell{c}{{Target SINR [dB]}} & 0 & 2 &
4 & 6 & 8 \\
\hline
Coordinated Beamforming & 72.80 & 72.80 & 72.80 & 72.80 & 72.80  \\
\hline
\tabincell{c}{$\ell_1/\ell_\infty$-Norm  Algorithm} & 36.08 & 43.76 & 52.36 &
60.16 & 69.56 \\
\hline
\tabincell{c}{Proposed Algorithm} & 30.40 & 38.08 & 45.56 &
56.76 & 70.48  \\
\hline
\end{tabular}
\label{rrh_large}
\end{table} 

\begin{table}[!t]
\renewcommand{\arraystretch}{1.3}
\caption{The Average Total Transmit Power Consumption with Different Algorithms
for Scenario Two}
\label{sparsity1}
\centering
\begin{tabular}{l|c|c|c|c|c}
\hline
\tabincell{c}{{Target SINR [dB]}} & 0 & 2 &
4 & 6 & 8 \\
\hline
Coordinated Beamforming & 3.02 & 5.16 & 8.84 & 15.05 & 25.41  \\
\hline
\tabincell{c}{$\ell_1/\ell_\infty$-Norm  Algorithm} & 8.54 & 10.96 & 14.43 &
19.87 & 27.42 \\
\hline
\tabincell{c}{Proposed Algorithm} & 8.03 & 11.25 & 16.32 &
20.03 & 26.28  \\
\hline
\end{tabular}
\label{transmit_large}
\end{table} 

We then consider a larger-sized network with  $L=8$ 2-antenna RRHs and $M=5$ multicast groups each has 2 single-antenna MUs, i.e., $|\Omega_m|=2,
\forall m$. The relative fronthaul links power consumption are set to be $P_l^c=[5.6+(l-1)]W, \forall l$. All error radii $\varepsilon_k$'s are set to be  $0.05$. Due to the high computational cost of the exhaustive search algorithm, we only simulate the $\ell_1/\ell_\infty$-norm based algorithm and the proposed robust group sparse beamforming algorithm. Fig. {\ref{large}}, Table {\ref{rrh_large}} and Table {\ref{transmit_large}} show the average network power consumption, the average relative fronthaul link power consumption and the average total transmit power consumption versus SINRs with different algorithms, respectively. From Fig. {\ref{large}}, we see that the proposed robust beamforming algorithm achieves lower network power consumption compared with the $\ell_1/\ell_\infty$-norm algorithm and the coordinated beamforming algorithm. In particular, Table {\ref{rrh_large}} shows that proposed algorithm achieves much lower relative fronthaul links power consumption, thought with a little higher transmit power consumption at the moderate target SINR regimes. Compared with the $\ell_1/\ell_\infty$-norm algorithm, the performance gain of the proposed algorithm is more prominent with low target SINRs. 

Overall, all the simulation results illustrate the effectiveness of the proposed robust group sparse beamforming algorithm to control  both the relative fronthaul  power consumption and the RRH transmit power consumption {\rev{with different network configurations}}.

\section{Conclusions and Future Works}
\label{confw}
This paper described a systematic way to develop computationally efficient algorithms based on the group-sparsity inducing penalty approach for the highly intractable network power minimization problem for multicast Cloud-RAN with imperfect CSI.  A novel quadratic  variational formulation of the weighted mixed $\ell_1/\ell_2$-norm was proposed to induce the group-sparsity structure for the robust multicast beamformer, thereby guiding the RRH selection. The perturbed alternating optimization, PhaseLift method, and SDR technique based algorithms were developed to solve the group-sparsity inducing optimization problem, the feasibility problems in RRH selection procedure and the transmit beamformer design problem in the final stage, respectively. Simulation results  illustrated the effectiveness of the proposed robust group sparse beamforming algorithm to minimize the network power consumption.  

Several future directions of interest are listed as follows:
\begin{itemize}
\item Although the proposed SDP based robust group sparse beamforming algorithm has a polynomial time complexity, the computational cost of the interior-point method will the prohibitive when the dimensions of the  SDP problems are large, such as in dense wireless networks. One may use the first-order method, e.g., the alternating direction method of multipliers (ADMM) \cite{boyd2011distributed, Boyd_arXiv2013, Yuanming_Globecom2014,Yuanming_LargeSOCP2014} to seek  modest accuracy solutions within reasonable time for the large-scale SDP problems \cite{Yuanming_ICC2015SDP}. 
\item It is desirable to lay the theoretical foundations for the tightness of the group-sparsity inducing penalty approach for finding approximate solutions to  the network power minimization problem as a mixed-integer non-linear optimization problem, and also for the tightness of PhaseLift method and SDR technique. 
\item It is interesting to apply the sparsity modeling framework to more mixed-integer nonlinear optimization problems, i.e., the joint user scheduling or admission and beamforming problems, which are essentially required to  control the sparsity structure and the magnitude of the beamforming coefficients.       
\end{itemize}

\appendices
\section{Proof of Theorem {\ref{conal}} }
\label{conprof}
We first consider problem $\mathscr{P}_{\textrm{GS}}(\epsilon)$ with a fixed $\epsilon$. Based on \cite[Theorem 4.9]{Gorski2007biconvex}, we know that the accumulation point $({\bs{\mu}}^{\star}(\epsilon), {\bf{Q}}^{\star}(\epsilon), {\bs{\lambda}}^{\star}(\epsilon))$ of the sequence $\{{\bs{\mu}}^{[i]}(\epsilon), {\bf{Q}}^{[i]}(\epsilon), {\bs{\lambda}}^{[i]}(\epsilon)\}$ converges to a stationary point of  problem $\mathscr{P}_{\textrm{GS}}(\epsilon)$, provided that the optimal solution (\ref{simplexup}) is unique with ${\bf{Q}}={\bf{Q}}^{\star}({\epsilon})$. This can be easily justified by the strict convexity of $\mathcal{R}_{\epsilon}({\bs{\mu}}, {\bf{Q}})$  with respect to ${\bs{\mu}}$ for a fixed ${\bf{Q}}$.

Next, we will prove the relationship (\ref{epscon})  between $\Lambda_{0}$ and $\Lambda(\epsilon)$. For convenience, we define the feasible region of problems $\mathscr{P}_{\textrm{GS}}$ and $\mathscr{P}_{\textrm{GS}}(\epsilon)$ as $\mathcal{C}$. Then problem $\mathscr{P}_{\textrm{GS}}(\epsilon)$ can be rewritten as
\begin{eqnarray}
\mathscr{P}_{\textrm{GS}}(\epsilon): 
\mathop {\textrm{minimize}}_{{\bf{x}}\in\mathcal{C}}&&
\mathcal{R}_{\epsilon}({\bf{x}}),
\end{eqnarray}
where ${\bf{x}}=({\bs{\mu}}, {\bf{Q}}, {\bs{\lambda}})$. Let $\Lambda_{0}$ and $\Lambda(\epsilon)$ denote the sets of the stationary points (or Karush-Kuhn-Tucker (KKT) pairs) of problems $\mathscr{P}_{\textrm{GS}}$ and $\mathscr{P}_{\textrm{GS}}(\epsilon)$ as
\begin{eqnarray}
\label{sta1}
\Lambda_{0}=\{{\bf{x}}\in\mathcal{C}: -\nabla_{\bf{x}}\mathcal{R}({\bf{x}})\in
\mathcal{N}_{\mathcal{C}}({\bf{x}})\},
\end{eqnarray}
and
\begin{eqnarray}
\label{sta2}
\Lambda(\epsilon)=\{{\bf{x}}\in\mathcal{C}: -\nabla_{\bf{x}}\mathcal{R}_{\epsilon}({\bf{x}})\in
\mathcal{N}_{\mathcal{C}}({\bf{x}})\},
\end{eqnarray}
respectively, where ${\mathcal{N}}_{\mathcal{C}}({\bf{x}})$ is the
normal cone \cite{rockafellar1998variational} to the convex set $\mathcal{C}$ at ${\bf{x}}$, i.e.,
\begin{eqnarray}
\mathcal{N}_{\mathcal{C}}({\bf{x}})=\{{\bf{v}}|\langle{\bf{v}}, {\bf{y}}-{\bf{x}}\rangle\le
0, \forall {\bf{y}}\in\mathcal{C}\}.
\end{eqnarray}
We first prove that 
\begin{eqnarray}
\label{proofweak}
\lim\sup_{\epsilon\searrow 0} \Lambda(\epsilon)\subset\Lambda_{0}.
\end{eqnarray}
Assuming that for any ${\bf{x}}^{\star}\in\lim\sup_{\epsilon\searrow 0}\Lambda(\epsilon)$, there exists $\epsilon_k\searrow 0$ and  ${\bf{x}}_{k}\in\Lambda(\epsilon_k)$ such that ${\bf{x}}_k\rightarrow {\bf{x}}^{\star}$. Based on \cite[Proposition 6.6]{rockafellar1998variational}, we have that
\begin{eqnarray}
\limsup_{{\bf{x}}_k\rightarrow {\bf{x}}^{\star}}\mathcal{N}_{\mathcal{C}}({\bf{x}}_k)=\mathcal{N}_{\mathcal{C}}({\bf{x}}^{\star}).
\end{eqnarray}
Furthermore, we have
\begin{eqnarray}
\label{prooflim}
-\nabla_{\bf{x}}\mathcal{R}_{\epsilon_k}({\bf{x}}_k)\in \mathcal{N}_{\mathcal{C}}({\bf{x}}_k),
\end{eqnarray}
and
\begin{eqnarray}
\lim_{k\rightarrow+\infty}\nabla_{\bf{x}}\mathcal{R}_{\epsilon_k}({\bf{x}}_k)&=&\lim_{\epsilon_k\searrow 0}\lim_{{\bf{x}}_{k}\rightarrow {\bf{x}}^{\star}}\nabla_{\bf{x}}\mathcal{R}_{\epsilon_k}({\bf{x}}_k)\nonumber\\
&=&\lim_{\epsilon_k\searrow 0}\nabla_{\bf{x}}\mathcal{R}_{\epsilon_k}({\bf{x}}^{\star})\nonumber\\
&=&\nabla_{\bf{x}}\mathcal{R}({\bf{x}}^{\star}).
\end{eqnarray}  
Therefore, taking $k\rightarrow +\infty$ in equation (\ref{prooflim}), we obtain that ${\bf{x}}^{\star}\in\Lambda_{0}$. We thus complete the proof for (\ref{proofweak}).

Define the {deviation}
of a given set ${A}_1$ from another set ${A}_2$ as \cite{shapiro2009lectures}
\begin{eqnarray}
\label{dist}
\mathbb{D}({A}_1, {A}_2)=\sup_{x_1\in {A}_1}\left(\inf_{x_2\in
{A}_2}\|x_1-x_2\|\right).
\end{eqnarray}
Based on the conclusion (\ref{proofweak}) and \cite[Theorem 4]{hong2011sequential}, we complete the proof for the conclusion  (\ref{epscon}). 
 
\section*{Acknowledgment}

The authors would like to thank Prof. Francis Bach for introducing reference \cite{Bach_arXiv2012convex}, which inspired the group spare beamforming modeling framework in Section {\ref{GSBFModel}}.
\bibliographystyle{ieeetr}

\begin{thebibliography}{10}

\bibitem{Foschini_2006network}
M.~K. Karakayali, G.~J. Foschini, and R.~A. Valenzuela, ``Network coordination
  for spectrally efficient communications in cellular systems,'' {\em IEEE
  Wireless Commun.}, vol.~13, pp.~56--61, Aug. 2006.

\bibitem{Jun_2009networkedTWC}
J.~Zhang, R.~Chen, J.~G. Andrews, A.~Ghosh, and R.~W. Heath, ``Networked {MIMO}
  with clustered linear precoding,'' {\em IEEE Trans. Wireless Commun.},
  vol.~8, pp.~1910--1921, Apr. 2009.

\bibitem{Gesbert_JSAC10}
D.~Gesbert, S.~Hanly, H.~Huang, S.~Shamai~Shitz, O.~Simeone, and W.~Yu,
  ``Multi-cell {MIMO} cooperative networks: A new look at interference,'' {\em
  IEEE J. Sel. Areas Commun.}, vol.~28, pp.~1380--1408, Sep. 2010.

\bibitem{mobile2011c}
{China Mobile}, ``C-{RAN}: the road towards green {RAN},'' {\em White Paper,
  ver. 3.0}, Dec. 2013.

\bibitem{Yuanming_TWC2014}
Y.~Shi, J.~Zhang, and K.~B. Letaief, ``Group sparse beamforming for green
  {C}loud-{RAN},'' {\em IEEE Trans. Wireless Commun.}, vol.~13, pp.~2809--2823,
  May 2014.

\bibitem{Yuanming_WCMLargeCVX}
Y.~Shi, J.~Zhang, K.~B. Letaief, B.~Bai, and W.~Chen, ``Large-scale convex
  optimization for ultra-dense {C}loud-{RAN},'' {\em IEEE Wireless Commun.
  Mag., to appear}, 2015.

\bibitem{Shamai_TSP2013}
S.-H. Park, O.~Simeone, O.~Sahin, and S.~Shamai, ``Joint precoding and
  multivariate backhaul compression for the downlink of cloud radio access
  networks,'' {\em IEEE Trans. Signal Process.}, vol.~61, pp.~5646--5658, Nov.
  2013.

\bibitem{Yuaning_ICC2014}
Y.~Shi, J.~Zhang, and K.~Letaief, ``{CSI} overhead reduction with stochastic
  beamforming for cloud radio access networks,'' in {\em Proc. of IEEE Int.
  Conf. on Commun. (ICC), Sydney, Australia}, Jun. 2014.

\bibitem{Yuanming_TSP14SCB}
Y.~Shi, J.~Zhang, and K.~Letaief, ``Optimal stochastic coordinated beamforming
  for wireless cooperative networks with {CSI} uncertainty,'' {\em IEEE Trans.
  Signal Process.}, vol.~63, pp.~960--973, Feb. 2015.

\bibitem{Jeff_JSAC5G}
J.~Andrews, S.~Buzzi, W.~Choi, S.~Hanly, A.~Lozano, A.~Soong, and J.~Zhang,
  ``What will 5{G} be?,'' {\em IEEE J. Sel. Areas Commun.}, vol.~32,
  pp.~1065--1082, Jun. 2014.

\bibitem{WeiYu_WC10}
H.~Dahrouj and W.~Yu, ``Coordinated beamforming for the multicell multi-antenna
  wireless system,'' {\em IEEE Trans. Wireless Commun.}, vol.~9,
  pp.~1748--1759, Sep. 2010.

\bibitem{Goldsmith_JSAC2004energy}
S.~Cui, A.~J. Goldsmith, and A.~Bahai, ``Energy-efficiency of mimo and
  cooperative mimo techniques in sensor networks,'' {\em IEEE J. Sel. Areas
  Commun.}, vol.~22, pp.~1089--1098, Aug. 2004.

\bibitem{love2008overview}
D.~J. Love, R.~W. Heath, V.~K. Lau, D.~Gesbert, B.~D. Rao, and M.~Andrews, ``An
  overview of limited feedback in wireless communication systems,'' {\em IEEE
  J. Sel. Areas Commun.}, vol.~26, pp.~1341--1365, Oct. 2008.

\bibitem{Jindal_TC2010unified}
N.~Jindal and A.~Lozano, ``A unified treatment of optimum pilot overhead in
  multipath fading channels,'' {\em IEEE Trans. Commun.}, vol.~58,
  pp.~2939--2948, Oct. 2010.

\bibitem{Tse_TIT2012completely}
M.~A. Maddah-Ali and D.~Tse, ``Completely stale transmitter channel state
  information is still very useful,'' {\em IEEE Trans. Inf. Theory}, vol.~58,
  pp.~4418--4431, Jul. 2012.

\bibitem{Jun_2009mode}
J.~Zhang, R.~W. Heath, M.~Kountouris, and J.~G. Andrews, ``Mode switching for
  the multi-antenna broadcast channel based on delay and channel
  quantization,'' {\em EURASIP J. Adv. Signal Process. (Special Issue Multiuser
  Lim. Feedback)}, vol.~2009, Article ID 802548, 15 pages, 2009.

\bibitem{Schaefer_SPM2014Physcial}
R.~Schaefer and H.~Boche, ``Physical layer service integration in wireless
  networks: Signal processing challenges,'' {\em IEEE Signal Process. Mag.},
  vol.~31, pp.~147--156, May 2014.

\bibitem{shamai2006capacity}
H.~Weingarten, Y.~Steinberg, and S.~Shamai, ``The capacity region of the
  gaussian multiple-input multiple-output broadcast channel,'' {\em IEEE Trans.
  Inf. Theory}, vol.~52, pp.~3936--3964, Sep. 2006.

\bibitem{Luo_2008quality}
E.~Karipidis, N.~D. Sidiropoulos, and Z.-Q. Luo, ``Quality of service and
  max-min fair transmit beamforming to multiple cochannel multicast groups,''
  {\em IEEE Trans. Signal Process.}, vol.~56, pp.~1268--1279, Mar. 2008.

\bibitem{Caire_CM2013}
N.~Golrezaei, A.~Molisch, A.~Dimakis, and G.~Caire, ``Femtocaching and
  device-to-device collaboration: A new architecture for wireless video
  distribution,'' {\em IEEE Commun. Mag.}, vol.~51, pp.~142--149, Apr. 2013.

\bibitem{bertsimas2011theory}
D.~Bertsimas, D.~B. Brown, and C.~Caramanis, ``Theory and applications of
  robust optimization,'' {\em SIAM review}, vol.~53, pp.~464--501, Aug. 2011.

\bibitem{Shamai_TIT2009multicast}
H.~Weingarten, T.~Liu, S.~Shamai, Y.~Steinberg, and P.~Viswanath, ``The
  capacity region of the degraded multiple-input multiple-output compound
  broadcast channel,'' {\em IEEE Trans. Inf. Theory}, vol.~55, pp.~5011--5023,
  Nov. 2009.

\bibitem{Jorswieck_TSP2011optimal}
R.~Mochaourab and E.~Jorswieck, ``Optimal beamforming in interference networks
  with perfect local channel information,'' {\em IEEE Trans. Signal Process.},
  vol.~59, pp.~1128--1141, Mar. 2011.

\bibitem{shapiro2009lectures}
A.~Shapiro, D.~Dentcheva, and A.~P. Ruszczy{\'n}ski, {\em Lectures on
  stochastic programming: modeling and theory}, vol.~9.
\newblock SIAM, 2009.

\bibitem{ben2009robust}
A.~Ben-Tal, L.~El~Ghaoui, and A.~Nemirovski, {\em Robust optimization}.
\newblock Princeton University Press, 2009.

\bibitem{shen2012distributed}
C.~Shen, T.-H. Chang, K.-Y. Wang, Z.~Qiu, and C.-Y. Chi, ``Distributed robust
  multicell coordinated beamforming with imperfect {CSI}: an {ADMM} approach,''
  {\em IEEE Trans. Signal Process.}, vol.~60, pp.~2988--3003, Jun. 2012.

\bibitem{Z.Q.Luo_SPM2010}
Z.-Q. Luo, W.-K. Ma, A.-C. So, Y.~Ye, and S.~Zhang, ``Semidefinite relaxation
  of quadratic optimization problems,'' {\em IEEE Signal Process. Mag.},
  vol.~27, pp.~20--34, May 2010.

\bibitem{boyd2004convex}
S.~P. Boyd and L.~Vandenberghe, {\em Convex optimization}.
\newblock Cambridge university press, 2004.

\bibitem{Z.Q.Luo_2012robust}
E.~Song, Q.~Shi, M.~Sanjabi, R.-Y. Sun, and Z.-Q. Luo, ``Robust
  {SINR}-constrained {MISO} downlink beamforming: When is semidefinite
  programming relaxation tight?,'' {\em EURASIP J. Wireless Communun. Netw.},
  vol.~2012, no.~1, pp.~1--11, 2012.

\bibitem{Bach_ML2011}
F.~Bach, R.~Jenatton, J.~Mairal, and G.~Obozinski, ``Optimization with
  sparsity-inducing penalties,'' {\em Foundations Trends Mach. Learning},
  vol.~4, pp.~1--106, Jan. 2012.

\bibitem{Z.Q.Luo_JSAC2013}
M.~Hong, R.~Sun, H.~Baligh, and Z.-Q. Luo, ``Joint base station clustering and
  beamformer design for partial coordinated transmission in heterogeneous
  networks,'' {\em IEEE J. Sel. Areas Commun.}, vol.~31, pp.~226--240, Feb.
  2013.

\bibitem{Mehanna_SP2013}
O.~Mehanna, N.~Sidiropoulos, and G.~Giannakis, ``Joint multicast beamforming
  and antenna selection,'' {\em IEEE Trans. Signal Process.}, vol.~61,
  pp.~2660--2674, May 2013.

\bibitem{Emil_TSP2012}
E.~Bj{\"o}rnson, G.~Zheng, M.~Bengtsson, and B.~Ottersten, ``Robust monotonic
  optimization framework for multicell {MISO} systems,'' {\em IEEE Trans.
  Signal Process.}, vol.~60, pp.~2508--2523, May 2012.

\bibitem{Bjornson_TCIT2013}
E.~Bj{\"o}rnson and E.~Jorswieck, ``Optimal resource allocation in coordinated
  multi-cell systems,'' {\em Found. Trends Commun. Inf. Theory}, vol.~9,
  pp.~113--381, Jan. 2013.

\bibitem{candes_2012phaseretrieval}
E.~J. Candes and X.~Li, ``Solving quadratic equations via phaselift when there
  are about as many equations as unknowns,'' {\em Found. Comput. Math.},
  pp.~1--10, Jun. 2012.

\bibitem{leyffer_2012mixed}
S.~Leyffer, {\em Mixed integer nonlinear programming}, vol.~154.
\newblock Springer, 2012.

\bibitem{Cheng_SP2013}
Y.~Cheng, M.~Pesavento, and A.~Philipp, ``Joint network optimization and
  downlink beamforming for {C}o{MP} transmissions using mixed integer conic
  programming,'' {\em IEEE Trans. Signal Process.}, vol.~61, pp.~3972--3987,
  Aug. 2013.

\bibitem{Yuanming_Globecom2013}
Y.~Shi, J.~Zhang, and K.~Letaief, ``Group sparse beamforming for green cloud
  radio access networks,'' in {\em Proc. IEEE Global Communications Conf.
  (GLOBECOM)}, pp.~4635--4640, Atlanta, GA, USA, Dec. 2013.

\bibitem{Bach_ML2008}
A.~Rakotomamonjy, F.~Bach, S.~Canu, and Y.~Grandvalet, ``Simple{MKL},'' {\em
  Journal of Machine Learning Research}, vol.~9, pp.~2491--2521, 2008.

\bibitem{Gorski2007biconvex}
J.~Gorski, F.~Pfeuffer, and K.~Klamroth, ``Biconvex sets and optimization with
  biconvex functions: a survey and extensions,'' {\em Mathematical Methods of
  Operations Research}, vol.~66, no.~3, pp.~373--407, 2007.

\bibitem{Argyriou_2008convexmutitask}
A.~Argyriou, T.~Evgeniou, and M.~Pontil, ``Convex multi-task feature
  learning,'' {\em Machine Learning}, vol.~73, no.~3, pp.~243--272, 2008.

\bibitem{Candes2013phase}
E.~J. Candes, Y.~C. Eldar, T.~Strohmer, and V.~Voroninski, ``Phase retrieval
  via matrix completion,'' {\em SIAM Journal on Imaging Sciences}, vol.~6,
  no.~1, pp.~199--225, 2013.

\bibitem{boyd2011distributed}
S.~Boyd, N.~Parikh, E.~Chu, B.~Peleato, and J.~Eckstein, ``Distributed
  optimization and statistical learning via the alternating direction method of
  multipliers,'' {\em Found. Trends in Mach. Learn.}, vol.~3, pp.~1--122, Jul.
  2011.

\bibitem{Boyd_arXiv2013}
B.~O'Donoghue, E.~Chu, N.~Parikh, and S.~Boyd, ``Conic optimization via
  operator splitting and homogeneous self-dual embedding,'' {\em arXiv preprint
  arXiv:1312.3039}, 2013, [Online]. Available: http://arxiv.org/abs/1312.3039

\bibitem{Yuanming_Globecom2014}
Y.~Shi, J.~Zhang, and K.~B. Letaief, ``Scalable coordinated beamforming for
  dense wireless cooperative networks,'' in {\em Proc. IEEE Global
  Communications Conf. (GLOBECOM)}, Austin, TX, 2014.

\bibitem{Yuanming_LargeSOCP2014}
Y.~Shi, J.~Zhang, B.~O'Donoghue, and K.~B. Letaief, ``Large-scale convex
  optimization for dense wireless cooperative networks,'' {\em IEEE Trans.
  Signal Process., to appear}, 2015.

\bibitem{Yuanming_ICC2015SDP}
J.~Cheng, Y.~Shi, B.~Bai, W.~Chen, J.~Zhang, and K.~Letaief, ``Group sparse
  beamforming for multicast green {C}loud-{RAN} via parallel semidefinite
  programming,'' {\em IEEE Int. Conf. on Commun. (ICC), London, UK}, 2015.

\bibitem{rockafellar1998variational}
R.~T. Rockafellar and R.~J.-B. Wets, {\em Variational analysis}, vol.~317.
\newblock Springer, 1998.

\bibitem{hong2011sequential}
L.~J. Hong, Y.~Yang, and L.~Zhang, ``Sequential convex approximations to joint
  chance constrained programs: A {M}onte {C}arlo approach,'' {\em Oper. Res.},
  vol.~59, pp.~617--630, May-Jun. 2011.

\bibitem{Bach_arXiv2012convex}
G.~Obozinski and F.~Bach, ``Convex relaxation for combinatorial penalties,''
  {\em arXiv preprint arXiv:1205.1240}, 2012, [Online]. Available: http://arxiv.org/abs/1205.1240

\end{thebibliography}

\end{document}